\documentclass[fleqn,usenatbib]{mnras}

\usepackage{newtxtext,newtxmath}
\usepackage[T1]{fontenc}
\usepackage{caption}
\usepackage{floatrow}
\DeclareNewFloatType{algorithm}{name=\textbf{Algorithm}}
\DeclareRobustCommand{\VAN}[3]{#2}
\let\VANthebibliography\thebibliography
\def\thebibliography{\DeclareRobustCommand{\VAN}[3]{##3}\VANthebibliography}

\usepackage{graphicx}	% Including figure files
\usepackage{amsmath}	% Advanced maths commands
\usepackage{amssymb}	% Extra maths symbols

\usepackage[dvipsnames]{xcolor}     % colorful comments 
\usepackage{ulem}
\usepackage{hyperref}
\usepackage{multirow}
\newcommand{\mpch}{{{\rm Mpc}~h^{-1}}}
\title[Self-calibration of interloper bias]{Precise self-calibration of interloper bias in spectroscopic surveys
}

\author[H. Peng and Y. Yu]{
Hui Peng,$^{1,2}$
and
Yu Yu$^{1,2}$\thanks{E-mail:~\href{mailto:yuyu22@sjtu.edu.cn}{yuyu22@sjtu.edu.cn}}
\\
$^{1}$Department of Astronomy, School of Physics and Astronomy, Shanghai Jiao Tong University, 800 Dongchuan Road, Shanghai 200240, China\\
$^{2}$Key Laboratory for Particle Astrophysics and Cosmology (MOE)/Shanghai Key Laboratory for Particle Physics and Cosmology, Shanghai 200240, China
}

\date{Accepted 2023 September 12. Received 2023 September 10; in original form 2023 May 16}

\pubyear{2023}

\begin{document}
\label{firstpage}
\pagerange{\pageref{firstpage}--\pageref{lastpage}}
\maketitle

\begin{abstract}
Interloper contamination due to line misidentification is an important issue in the future low-resolution spectroscopic surveys.
We realize that the algorithm previously used for photometric redshift self-calibration, with minor modifications, can be particularly applicable to calibrate the interloper bias.
In order to explore the robustness of the modified self-calibration algorithm, we construct the mock catalogues based on \textit{China Space Station Telescope} (\textit{CSST}), taking two main target emission lines, H$\alpha$ and [O\,\textsc{iii}].
The self-calibration algorithm is tested in cases with different interloper fractions at 1 per cent, 5 per cent and 10 per cent.
We find that the interloper fraction and mean redshift in each redshift bin can be successfully reconstructed at the level of $\sim$~0.002 and $\sim$~$0.001(1+z)$, respectively.
We also find the impact of the cosmic magnification can be significant, which is usually ignored in previous works, and therefore propose a convenient and efficient method to eliminate it.
Using the elimination method, we show that the calibration accuracy can be effectively recovered with slightly larger uncertainty.
\end{abstract}


%%%%%%%%%%%%%%%%%%%%%%%%%%%%%%%%%%%%%%%%%%%%%%%%%%

%%%%%%%%%%%%%%%%% BODY OF PAPER %%%%%%%%%%%%%%%%%%

\section{Introduction}
\label{sec:intro}
Spectroscopic surveys can be very powerful to explore the formation and evolution of galaxies and the large-scale structure (LSS), and study the properties of dark matter and dark energy.
There are several ongoing and upcoming spectroscopic surveys to observe larger and deeper maps of sky, such as the \textit{Nancy Grace Roman Space Telescope}\footnote{\href{https://roman.gsfc.nasa.gov}{https://roman.gsfc.nasa.gov}} \citep[\textit{RST};][]{Spergel:2015aa}, the Prime Focus Spectrograph\footnote{\href{https://pfs.ipmu.jp}{https://pfs.ipmu.jp}} \citep[PFS;][]{Takada:2014vd}, the Hobby–Eberly Telescope Dark Energy Experiment\footnote{\href{https://hetdex.org}{https://hetdex.org}} \citep[HETDEX;][]{Gebhardt:2021aa}, the \textit{Euclid}\footnote{\href{https://www.cosmos.esa.int/web/euclid}{https://www.cosmos.esa.int/web/euclid}} \citep{Amendola:2018aa}, the Dark Energy Spectroscopic Instrument\footnote{\href{https://www.desi.lbl.gov}{https://www.desi.lbl.gov}} \citep[DESI;][]{DESI-Collaboration:2016vy,DESI-Collaboration:2016vs}, and the \textit{China Space Station Telescope} \citep[\textit{CSST};][]{Gong:2019tb}.

Many of the spectroscopic surveys will obtain spectrum with lower signal-to-noise ratio ($\rm S/N$) to observe more galaxies at higher redshift, and the redshifts are likely determined by only a single emission line operating near the minimum acceptable $\rm S/N$ (e.g. HETDEX, \textit{RST}, \textit{Euclid}, and \textit{CSST}).
Thus, one important systematic effect is interloper contamination due to misidentified emission lines.
\citet{Pullen:2016vt} found that the interlopers can be eﬀectively removed in the surveys with high spectral resolution by secondary line identification or finding correlated emission lines, but should be seriously considered in the low-resolution spectroscopic surveys.
The \textit{RST} [O\,\textsc{iii}] survey may suffer from Ha contamination at a level of tens of percent, despite implementing secondary line identification, and this problem is expected to be more severe for \textit{Euclid} and \textit{CSST} due to their lower signal-to-noise line detection threshold \citep{Pullen:2016vt,Addison:2019aa}.
It has been shown that interlopers can significantly degrade the cosmological constraints from the measurements of power spectrum, correlation function, redshift-space distortions (RSD), baryon acoustic oscillations (BAO), weak lensing, and etc \citep[e.g.][]{Pullen:2016vt,Leung:2017aa,Addison:2019aa,Grasshorn-Gebhardt:2019aa,Awan:2020aa,Massara:2021aa}.

Several methods have been proposed to vanish the impact of interlopers, such as combination with photometry \citep{Kirby:2007aa} and Bayesian framework based on prior assumptions on the luminosity functions and equivalent width distributions of galaxies \citep{Leung:2017aa,Davis:2023aa,Mentuch-Cooper:2023aa}.
The measured correlation functions and power spectra are found to be significantly affected by the presence of interlopers and can be used to put constraints on it.
In \citet{Grasshorn-Gebhardt:2019aa}, they present a proof of concept for the constraints on the interloper fractions by including its effects in the modelling of the galaxy auto- and cross-power spectra of the main and the interloper samples.
An alternative approach is to minimize the residual difference between the observed cross-correlation multipoles and the predicted cross-correlation multipoles of the two samples \citep{Farrow:2021aa}.
Besides, \citet{Gong:2021tb} proposed that the galaxy-galaxy correlation between two particular observed redshift bins can be also explored to obtain the interloper fraction.
Similarly, a method is presented in \citet{Foroozan:2022aa} that fits a model for the monopole and quadrupole moments of the contaminated correlation function with a free parameter for the interloper fraction.

In this paper, we also focus on the galaxy-galaxy correlation between redshift bins, similar to the input used in \citet{Gong:2021tb}.
The straightforward approximate calculation with the ratio of angular correlation functions made in \citet{Gong:2021tb} results in a relatively large and unstable error in the reconstruction accuracy, especially for the cases with large interloper fractions.
We attempt to take one step further, to extend the interloper fraction estimates to a higher precision, meeting the requirements of the analyses in future spectroscopic surveys.
The theory of this method is almost the same as the one used in self-calibrating the photometric redshift errors \citep{Schneider:2006ta,Benjamin:2010aa,Zhang:2010wr,Zhang:2017um,Schaan:2020up,Peng:2022aa,Xu:2023aa}.
In both cases, the cross correlations between different observed redshift bins come from the redshift error.
Since the principle is the same, we find the algorithm previously used for photometric redshift self-calibration in \citet{Peng:2022aa} has enormous potential to solve this problem.

We propose the modified self-calibration algorithm to precisely obtain the interloper fractions in redshift bins of spectroscopic surveys.
To systematically investigate the performance of the algorithm in practice, we take the \textit{CSST} slitless spectroscopic survey as an example, and construct light-cone simulations to generate mock galaxy catalogs.
We take into account two main target lines in \textit{CSST}, H$\alpha$ and [O\,\textsc{iii}], and investigate on several interloper fraction cases.
We note that these two strong emission lines are also the most dominant targets used in \textit{RST} and \textit{Euclid}, with different redshift coverage due to the different filter design from \textit{CSST}.
In a wide interloper fraction range, we find that the algorithm can reconstruct accurate interloper fraction and the mean redshift in each tomographic bin.
We also find the cosmic magnification, which has been ignored in the past literature \citep[e.g.][]{Pullen:2016vt, Grasshorn-Gebhardt:2019aa,Farrow:2021aa,Gong:2021tb,Foroozan:2022aa}, however, can seriously contaminate the cross correlation measurements and degrade the accuracy.
Thus, we propose a convenient and efficient method to eliminate its impact.
The self-calibration results after implementing our elimination method are confirmed to be unbiased with very mild increase of the uncertainty.
Additionally, in Appendix \ref{appendix_B} we compliment the performance of our method on the contamination between H$\alpha$ and [O\,\textsc{ii}] emission lines.
We focus on \textit{CSST} here but the implications generalize to all similar low-resolution spectroscopic surveys.

This paper is organized as follows. In Section \ref{sec:method}, we give a brief overview of the self-calibration method and the modified algorithm.
The simulation and mock catalogues we use are presented in Section \ref{sec:simu}.
Section \ref{sec:results} describes the implementations of the algorithm and the method to eliminate the impact of the cosmic magnification.
Finally, we conclude and discuss in Section \ref{sec:conclusion}.
%%%%%%%%%%%%%
%%%%%%%%%%%%%
\section{Methodology}
\label{sec:method}

\subsection{The self-calibration method}
%\begin{itemize}
%    \item origin of the interloper bias
%    \item self-calibration method
%    \item specified redshift bin and example figure to show the scattering matrix
%\end{itemize}

\begin{figure*}
\centering
    \begin{minipage}{8.5cm}
        \centering
        \includegraphics[width=8.5cm]{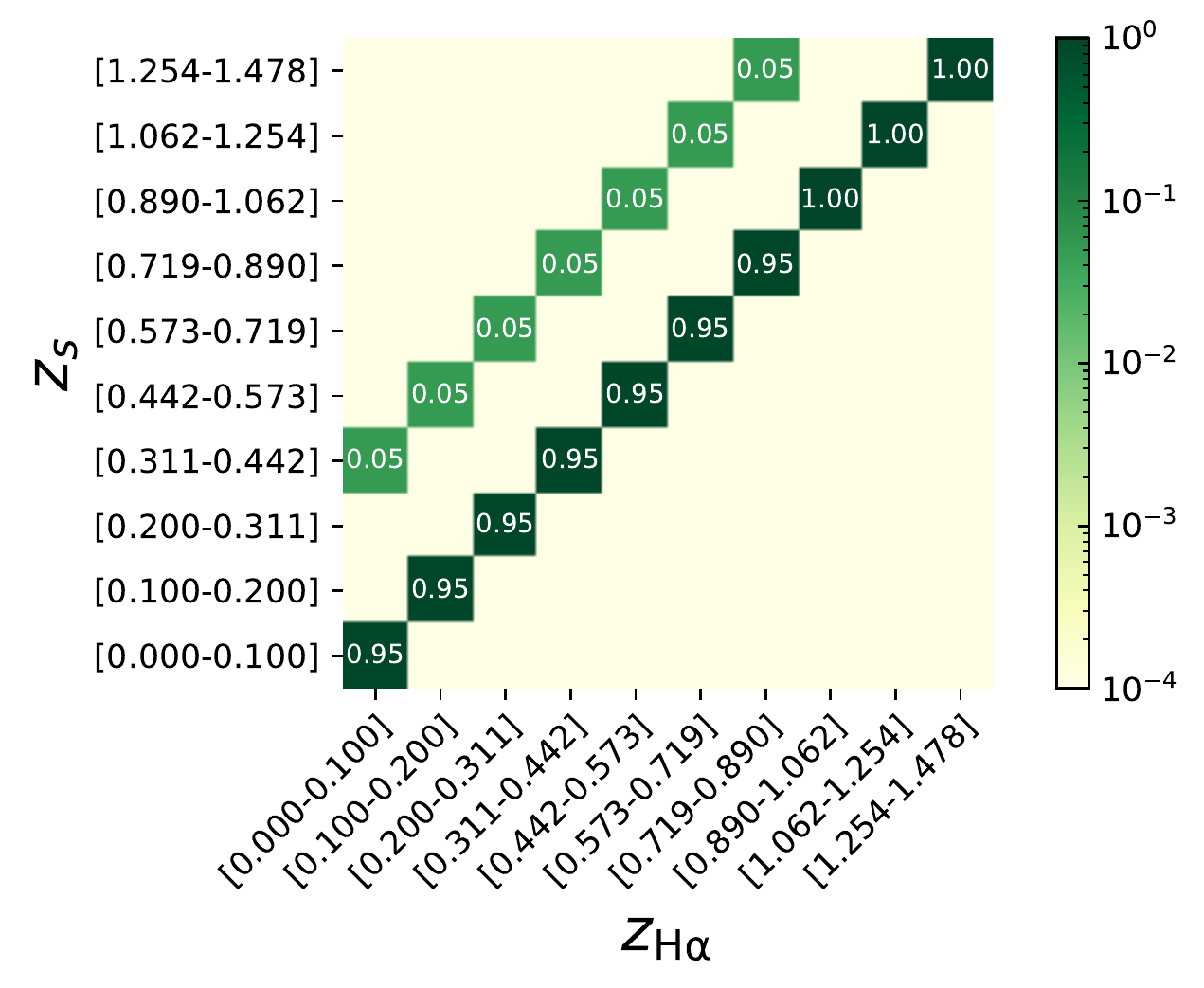}
    % \centerline{(a)}
    \end{minipage}
    % \hspace{1cm}
	\begin{minipage}{8.5cm}
    	\centering
        \includegraphics[width=8.5cm]{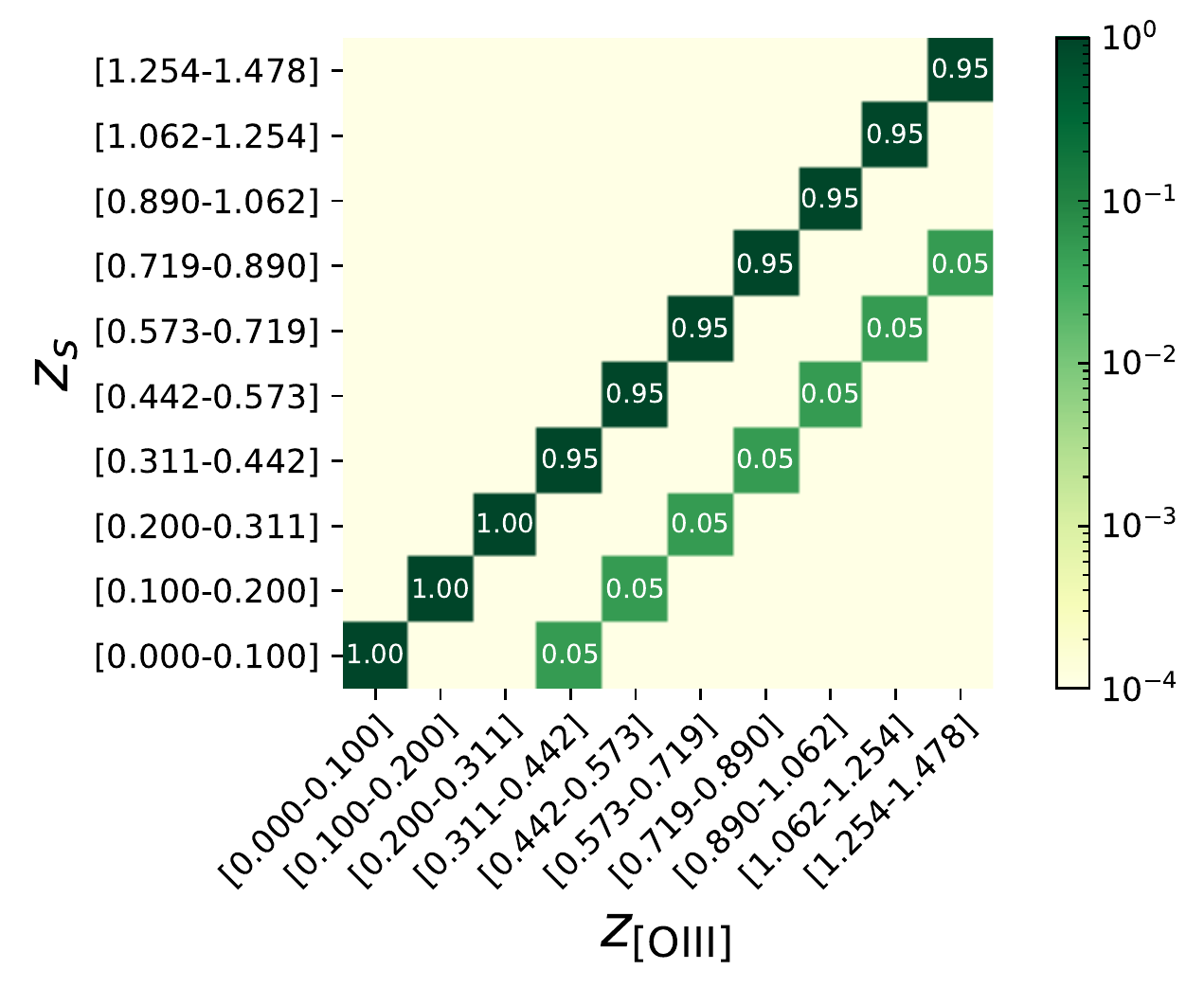}
    \end{minipage}
    \caption{Two typical cases when using H$\alpha$ (left panel) and [O\,\textsc{iii}] (right panel) lines to determine the redshifts with interloper fraction 5 per cent. The vertical axis $z_s$ labels the true redshift bins, the horizontal axis $z_{\rm H\alpha}$ and $z_{\rm [O\,\textsc{iii}]}$ denote the redshift bins detected through emission lines. The 10 redshift bins are divided according to the strict corresponding relation in equation~(\ref{eqn:interloper}).}
    \label{fig:example}
\end{figure*}

Assume that we observe the emission line at wavelength $\lambda_{\rm obs}$ in the spectrum.
It may come from one emission line with rest-frame wavelength $\lambda_1$ at redshift $z_1$ or another emission line with rest-frame wavelength $\lambda_2$ at a diﬀerent redshift $z_2$, which can be expressed as follows:
\begin{equation}
    \lambda_{\rm o b s}=\lambda_{1}\left(1+z_{1}\right)=\lambda_{2}\left(1+z_{2}\right)\ .
    \label{eqn:interloper}
\end{equation}
Thus, due to the possible misidentification, if we use $\lambda_1$ ($\lambda_2$) to determine the redshifts, there will be some interloper galaxies actually from $\lambda_2$ ($\lambda_1$) with different redshifts.
This interloper bias should be precisely calibrated to meet the requirements of the ongoing and future spectroscopic surveys.

The self-calibration method we use here is almost the same as the one used in calibrating the photometric redshift errors \citep{Peng:2022aa}.
Assume that we split galaxies into $n$ redshift bins.
We denote the ratio of the galaxies in true redshift bin $i$ but observed in redshift bin $j$ detected through an emission line as $P_{ij}\,{\equiv}\,N_{i{\rightarrow}j}/N_j^D$.
Here, $N_{i{\rightarrow}j}$ is the number of galaxies misidentified from the true-$z$ bin $i$ to the observed bin $j$, and $N_j^D$ is the total number of galaxies in the observed redshift bin $j$.
We have the normalization $\sum_iP_{ij}=1$.
The power spectrum of two observed redshift bins, $C_{ij}^{gg,D}(\ell)$, and the power spectrum of true redshift bins, $C_{ij}^{gg,R}(\ell)$, is related by
\begin{equation}
    C_{ij}^{gg, D}(\ell)=\sum_kP_{ki}P_{kj}C_{kk}^{gg,R}(\ell)+\delta{N_{ij}^{gg,D}(\ell)}\ .
    \label{eqn:CDsum}
\end{equation}
This equation has approximated $C_{k{\neq}m}^{gg,R}(\ell)=0$, as the galaxy cross-correlation between non-overlapping redshift bins would vanish under the Limber approximation and without the cosmic magnification.
The last term $\delta{N_{ij}^{gg,D}(\ell)}$ is the associated shot noise fluctuation after the subtraction of its ensemble average.
With the measurements on different scales, the number of unknowns ($P_{ij}$, $C_{kk}^{gg,R}$) will be less than the number of observables ($C_{ij}^{gg,D}$) and then the equation can be solved in principle.
For a given $\ell$, we can rewrite the above equations in matrix form,
\begin{equation}
    C_{\ell}^{gg,D}=P^{T}C_{\ell}^{gg,R}P+\delta{N_{\ell}^{gg,D}}\ .
    \label{eqn:CD}
\end{equation}

Note that there are some obvious difference between the interloper contamination in spectroscopic surveys and photometric redshift scattering in photometric surveys.
According to equation~(\ref{eqn:interloper}), there exist strict corresponding relation between the true redshifts and observed redshifts of interlopers when taking two emission lines into consideration.
If we divide the redshift bins properly, the interlopers will come from only one redshift bin, instead of from multiple bins, or even worse from all bins in the case of photometric redshift scattering.
Thus, there is only one non-zero $P_{i\ne j}$, and
we can denote this non-zero $P_{ij}$ as the interloper fraction $f_{\rm i}$ for the observed redshift bin $j$, and $P_{jj}$ is simply reduced to $1-f_{\rm i}$.
Take the spectroscopic survey like \textit{CSST} which observing H$\alpha$ 6563~{\AA} and [O\,\textsc{iii}] 5007~{\AA} lines as an example.
We can divide the redshift range $0<z<1.478$ into 10 tomographic bins with edges $z=$ 0.000, 0.100, 0.200, 0.311, 0.442, 0.573, 0.719, 0.890, 1.062, 1.254, 1.478.
The contamination happens between bin $j$ and $j+3$, for $j=1, \cdots, 7$.
% , i.e. (0.000-0.100), [0.100-0.200), [0.200-0.311), [0.311-0.442), [0.442-0.573), [0.573-0.719), [0.719-0.890), [0.890-1.062), [1.062-1.254) and [1.254-1.478).
% These 10 redshift bins are much narrower than the 3 redshift bins used in \citet{Gong:2021tb} and the maximum redshift is increased from 1.2 to 1.478, which can have more practical applications in future surveys.

Fig.~\ref{fig:example} shows two cases using these two emission lines to determine the redshifts with interloper fraction $f_{\rm i}=$ 5 per cent in each redshift bin.
We can easily figure out that the number of unknown parameters, $P_{ij}$, is greatly reduced, compared to the case of photometric redshift errors, making it possible to obtain precise solutions of the interloper fraction in each observed redshift bin.
In \citet{Gong:2021tb}, the ratio between the cross correlation and auto correlation is used to estimate the same interloper fractions here.
However, the accuracy of this simplistic approximation will degrade as the interloper fraction rises (though this can be mitigated in an iterative way, see Appendix \ref{appendix_A} for details).
Therefore, in order to assure the high accuracy for the future analysis in surveys, a more robust algorithm is needed to obtain the precise interloper fractions.

\subsection{Modified algorithm}
\label{sec:algorithm}
%\begin{itemize}
%    \item set 0 in corresponding matrix elements in iteration process of algorithm and cross power spectrum in data. 
%    \item do not need to limit the redshift width or cut on large scale.
%    \item only use results from Algorithm 1.
%    \item flow chart
%\end{itemize}
In \citet{Peng:2022aa}, an algorithm based on the self-calibration theory above was developed to calibrate the photometric redshift errors.
The technique attempts to find the result by minimizing
\begin{equation}
    \mathcal{J}\left(P ; C_{\ell=1, \ldots, N_{\ell}}^{g g, R}\right) \equiv \frac{1}{2} \sum_{\ell}\left\|C_{\ell}^{g g, D}-P^{T} C_{\ell}^{g g, R} P\right\|_{F}^{2}\ ,
    \label{define:J}
\end{equation}
where $\|.\|_F$ is the Frobenius form.
$\mathcal{J}$ measures the accumulation of decomposition error across all data matrices between the observations and reconstructions, where $C_{\ell}^{gg,D}$ is the observational power spectrum, $P$ and $C_{\ell}^{gg,R}$ are the derived results from the algorithm.
We find this self-calibration algorithm can also be very helpful to obtain the interloper fractions in observed redshift bins of spectroscopic surveys.
Of course, due to some unique properties in the cases of interloper bias, we implement the following modifications on the algorithm to make the reconstruction of interloper fractions more efficient.

%Firstly, according to equation~(\ref{eqn:interloper}) and Fig.~\ref{fig:example}, 
Firstly, the cross power spectra in $C_{\ell}^{g g, D}$ and the off-diagonal elements in matrix $P$ should only exist for the redshift bin pairs satisfying equation~(\ref{eqn:interloper}).
Thus, we set 0 for other matrix elements in iteration process of algorithm, and only the non-vanishing cross power spectra are used to initialize the iteration.
Secondly, for our redshift binning scheme, the interlopers will not come from adjacent redshift bin.
We do not need to worry about the large scale non-vanishing cross-correlation between neighbouring redshift bins when the bins are narrow.
Therefore, we relax the limitation on the largest scale and the redshift bin width used in the self-calibration for photometric redshift \citep{Peng:2022aa}.
%limit the redshift bin width or cut on large scale to make sure $C_{k{\neq}m}^{gg,R}(\ell)=0$ in equation~(\ref{eqn:CDsum}).
Finally, thanks to the extremely decreased number of unknown parameters and complexity of the system, here we only use the first part of the original algorithm for photometric redshift self-calibration, i.e. the fixed-point iteration algorithm, with the above modifications.
The new procedure is summarized in Algorithm \ref{tab:algorithm}.
We find that it is sufficient to solve the problem and obtain the accurate results.
We refer the readers to \citet{Zhang:2017um} and \citet{Peng:2022aa} for the original algorithm used in photometric redshift self-calibration.

\begin{algorithm}
	\caption{Modified algorithm for solving equation~(\ref{eqn:CD}) to obtain the interloper fractions in redshift bins. The differences from the algorithm used in \citet{Peng:2022aa} are shown in italics.}
	\label{tab:algorithm}
	\small
% 	\resizebox{\textwidth}{!}{
	\begin{tabular}{l}
		\hline
		\hline
		$\mathbf {Input}$: Matrices $C_{\ell}^{g g, D}$ for all $\ell$ (\textit{set 0 in corresponding elements}). \\
		$\mathbf {initialize}$: Assign $P$ to a random diagonal dominating initial matrix (\textit{set}\\
		\textit{0 in corresponding elements}) with $\sum_{i} P_{i j}=1$, for all $i$.\\
        $\mathbf {repeat}$\\
        for $\ell=1$ to $n_{\ell}$ do\\
        $C_{\ell}^{g g, R}=P^{T-1} C_{\ell}^{g g, D} P^{-1}$\\
        $Q_{\ell}=C_{\ell}^{g g, D} P^{-1}$\\
        $P^{T}={\rm A b s}[\left(\sum_{\ell} Q_{\ell}\right)(\sum_{\ell} C_{\ell}^{g g, R})^{-1}]$ (\textit{set 0 in corresponding elements})\\
        $v_{j}=\sum_{k}\left[P^{T}\right]_{j k}$, for all $j$\\
        $\left[P^{T}\right]_{i j}=\frac{\left[P^{T}\right]_{i j}+\epsilon}{v_{j}+\epsilon}$, for all $(i, j)$ (a typical value of $\epsilon$ is $10^{-10}$)\\
        $\mathbf {end~for}$\\
        $C_{\ell}^{g g, R}={\rm Abs[{Diag}}(C_{\ell}^{g g, R})]$, for all $\ell$\\
        until a truncation criterion or maximum step number is satisfied\\
        $\mathbf {Output}$: $P$\\
		\hline
	\end{tabular}
% 	}
\end{algorithm}

\section{Simulation}
\label{sec:simu}
%\begin{itemize}
%    \item zCOSMOS redshift distribution
%    \item simulation and mock
%\end{itemize}
\begin{figure}
    \centering
	\includegraphics[width=\columnwidth]{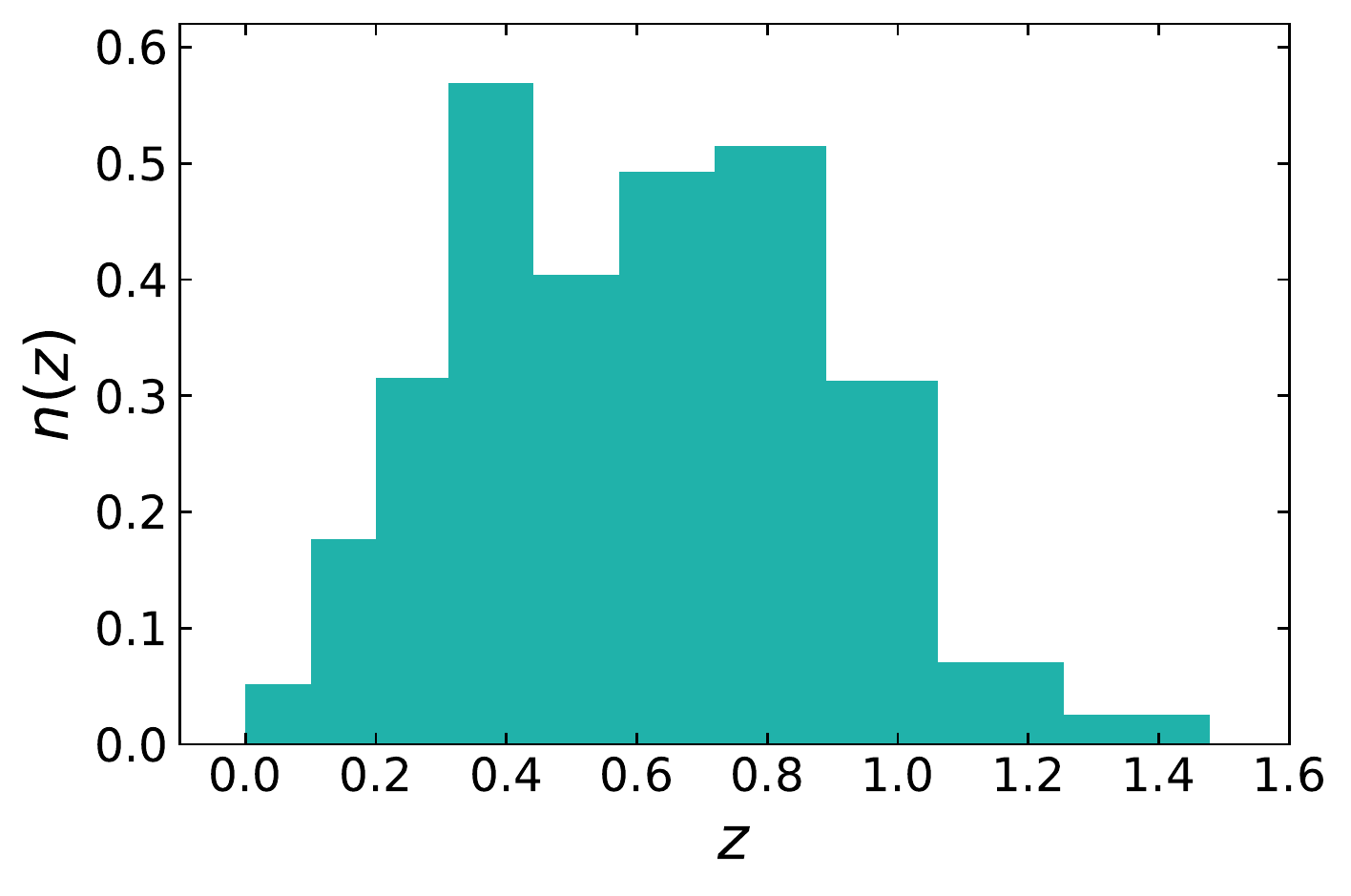}
    \caption{The redshift distribution of the zCOSMOS catalog, which is adopted to construct the mock catalogue for \textit{CSST} spectroscopic survey. The vertical axis indicates the number of objects per $\text{arcmin}^2$ in each observed redshift bin.}
    \label{fig:n_z}
\end{figure}

%The \textit{CSST} spectroscopic galaxy survey \citep{Gong:2019tb} is selected as an example to run simulations.
The \textit{CSST} is a 2~m space telescope which is planned to be launched at the end of 2024.
It will concurrently investigate both photometric imaging and slitless spectroscopic surveys, covering 17\,500~$\rm deg^2$ sky area in about 10 years.
The \textit{CSST} contains three bands, i.e. GU, GV and GI, with wavelength range from $\sim$~250~nm to $\sim$~1100~nm.
The H$\alpha$ and [O\,\textsc{iii}] emission lines are main targets in the \textit{CSST} spectroscopic survey.

The zCOSMOS is a survey covers 1.7~$\rm deg^2$ with a magnitude limit $\simeq22.5$ \citep{Lilly:2007wd,Lilly:2009vb}, closing to survey depth of the \textit{CSST}.
Thus, we use the redshift distribution of the zCOSMOS catalogue to construct the mock of \textit{CSST} spectroscopic survey, and the magnitude is limited to 22.5.
The redshift distribution is presented in Fig.~\ref{fig:n_z}, and the redshift bins are divided as in Fig.~\ref{fig:example}, ranging from 0 to 1.478.
Note that the redshift range covered by observing a specified emission line is limited by the wavelength coverage of instruments in different surveys, and can not reach 1.478 for H$\alpha$ or [O\,\textsc{iii}] in \textit{CSST} spectroscopic survey.
Here we use this broader redshift range from zCOSMOS as an example to show the universality of our method, without strict constrains on the redshift ranges of different emission lines in real spectroscopic surveys.

The simulation we use to construct mock galaxy catalogues is the same as in \citet{Peng:2022aa}, which is a high-resolution $N$-body simulation presented in \citet{Jing:2019uw} with a flat $\Lambda$CDM cosmological model consistent with the \textit{WMAP} observations \citep{Komatsu:2011un, Hinshaw:2013vg}.
By a particle-particle-particle-mesh gravity solver from $z=144$, the simulation evovles $3072^3$ particles inside a $(600~\mpch)^3$ comoving volume with periodical condition.
From the snapshots at various redshifts, we cut out curved slices with $300~\mpch$ thickness and stack them to construct light-cones upto $z$~$\sim$~2.48.
In order to prevent the repeating structures along line-of-sight, all the boxes are randomly rotated and shifted before slicing.
We made a total of 300 pseudo-independent light-cones, including lensing convergence maps, friends-of-friends haloes, and dark matter particle distributions.

We use 223 simulated maps, with 67.13~$\rm{deg}^2$ each, to cover $\sim$~15\,000~$\rm {deg^2}$,  similar to the sky map area of \textit{CSST} after masking.
To construct mock catalogues, we use the following steps.
Firstly, we set the number of galaxies in each observed bin using the redshift distribution in Fig.~\ref{fig:n_z}.
Then, according to the interloper fractions we preset, the number of galaxies in each true redshift bin can be fixed.
The halos generated in simulation are regarded as the galaxies we need.
We pick the galaxies in descending order of halo mass until the number is satisfied in each true redshift bin.
Finally, to match the interloper fractions, we randomly assign corresponding fraction of galaxies in true redshift bins to each observed bin.
Note that the observed redshifts of interlopers are known to us due to the strict relationship in equation~(\ref{eqn:interloper}).

The \textsc{healpix} \citep{Gorski:2005te} is used to construct galaxy overdensity map, with $N_{\text{side}}=1024$, corresponding to a spatial resolution of $\sim$~3.4~arcmin.
We use the function \textsc{compute\_coupled\_cell} in \textsc{namaster} \citep{Alonso:2019aa} to measure the angular power spectrum of galaxies.
As mentioned in Section \ref{sec:algorithm}, here we do not cut out the data on large scale.
Thus the $\ell$ modes from the fundamental frequency $\ell_{\rm{min}}=44$ to $\ell_{\rm{max}}=1000$ are used in the analysis and are divided into 6 broad bands, $[44, 410)$, $[410, 578)$, $[578, 708)$, $[708, 817)$, $[817, 913)$, $[913, 1000)$.

%%%%%%%%%%%%%
%%%%%%%%%%%%%
\section{Results}
\label{sec:results}
% \begin{table*}
% 	\centering
% 	\caption{The mean absolute bias.
% 	}
% 	\label{tab:result}
% 	\setlength{\tabcolsep}{1.25cm}{
% 	\begin{tabular}{ccccc}
% 		\hline
% 		\multirow{2}{*}{Cases}&\multicolumn{2}{c}{fiducial}&\multicolumn{2}{c}{eliminate the cosmic magnification}\\
% 		\cline{2-3}
% 		\cline{4-5}
% 		  &$\rm H\alpha$& [O\,\textsc{iii}] & $\rm H\alpha$& [O\,\textsc{iii}]\\
% 		 \hline
% 		$\langle f^{\rm recover}_{\rm i}\rangle-\langle f^{\rm true}_{\rm i}\rangle$ & 0.0014 & 0.0017 & ? & ?\\
% 		$\delta_{\langle z\rangle}/\rm (1+\langle z\rangle)$ & ? & ? & ? & ?\\
% 		\hline
% 	\end{tabular}}
% \end{table*}

% \begin{table}
% 	\centering
% 	\caption{The mean absolute bias of reconstructed interloper fraction.
% 	}
% 	\label{tab:result}
% 	\setlength{\tabcolsep}{0.58cm}{
% 	\begin{tabular}{ccc}
% 		\hline
% 		\multirow{2}{*}{Cases}&\multicolumn{2}{c}{$|\langle f^{\rm recover}_{\rm i}\rangle-\langle f^{\rm true}_{\rm i}\rangle|$}\\
% 		\cline{2-3}
% 		  &$\rm H\alpha$& [O\,\textsc{iii}]\\
% 		\hline
% 		fiducial&0.0017 & 0.0021\\
% 		eliminate cosmic magnification&? & ?\\
% 		\hline
% 	\end{tabular}}
% \end{table}
\subsection{Fiducial results}
\label{sec:fiducial}

\begin{figure*}
\centering
    \begin{minipage}{8cm}
        \centering
        \includegraphics[width=8cm]{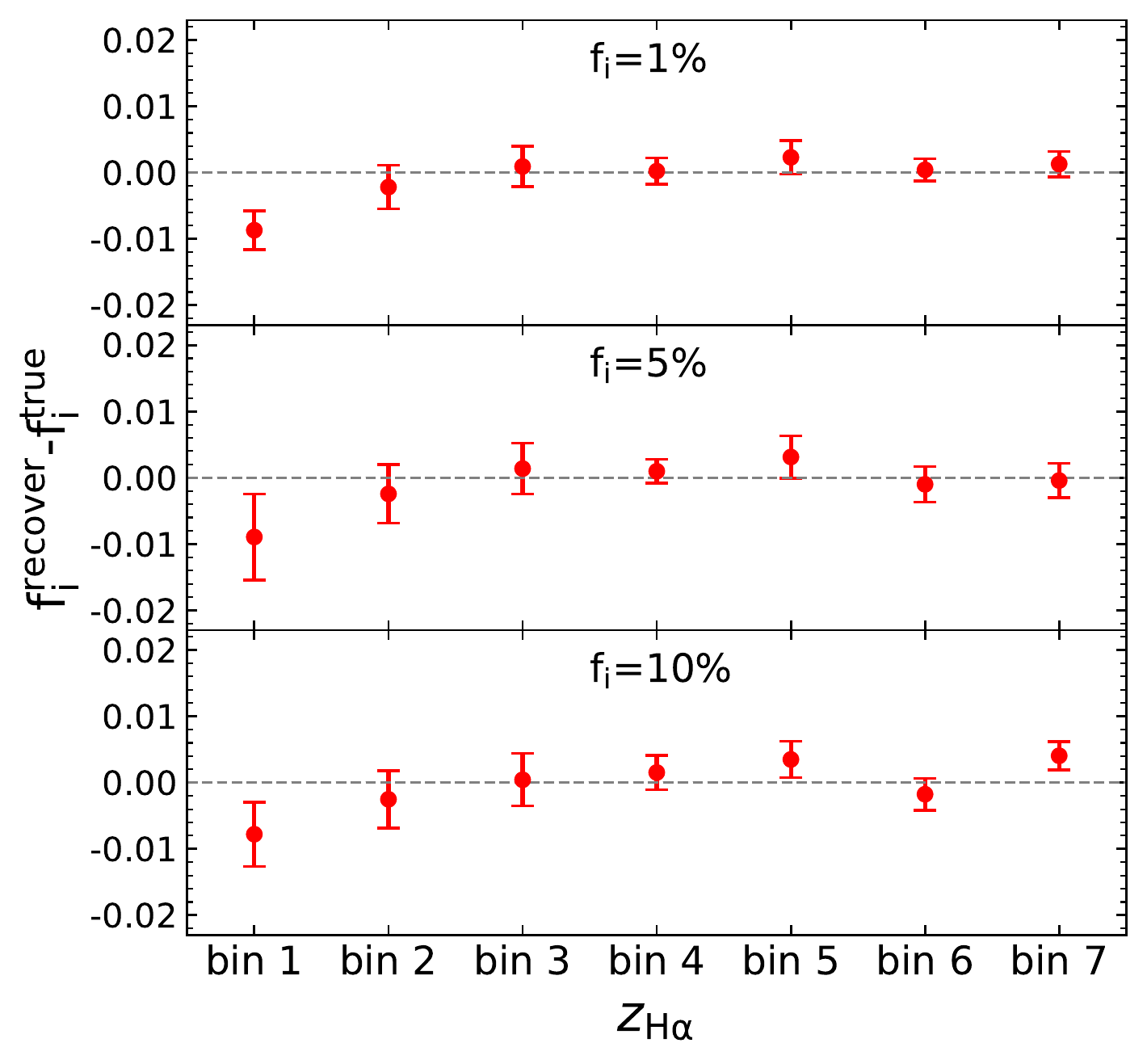}
    \end{minipage}
    \hspace{0.3cm}
	\begin{minipage}{8cm}
    	\centering
        \includegraphics[width=8cm]{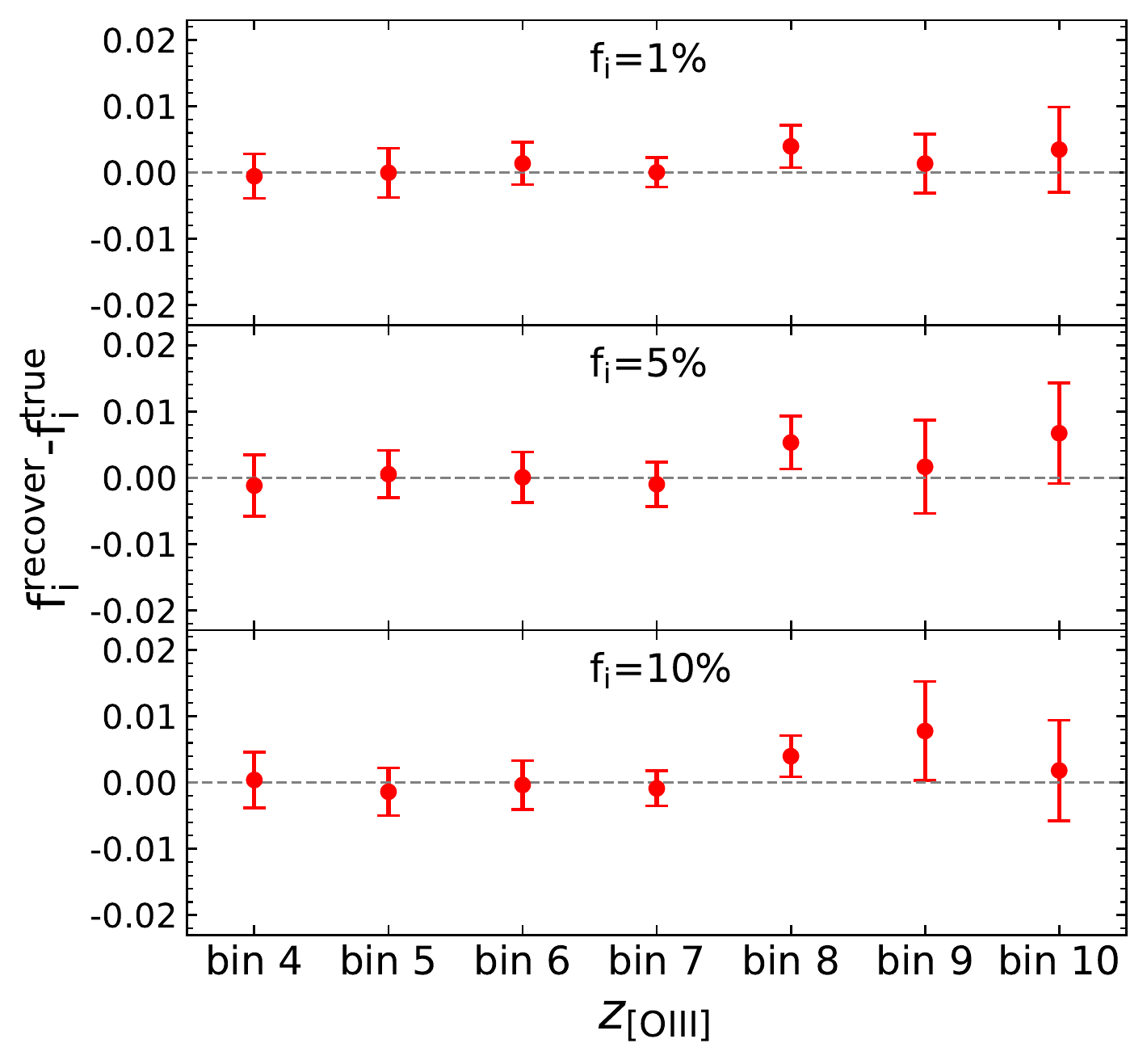}
    \end{minipage}
    \caption{The reconstruction results when using $\rm H\alpha$ (left panel) and $\rm [O\,\textsc{iii}]$ (right panel) lines to determine the redshifts of galaxies, with interloper fractions at 1 per cent, 5 per cent and 10 per cent. The points and error bars indicate the biases and $1\sigma$ uncertainties of the reconstructed interloper fractions in different cases.}
    \label{fig:result_interloper}
\end{figure*}

\begin{figure*}
\centering
    \begin{minipage}{8.5cm}
        \centering
        \includegraphics[width=8.5cm]{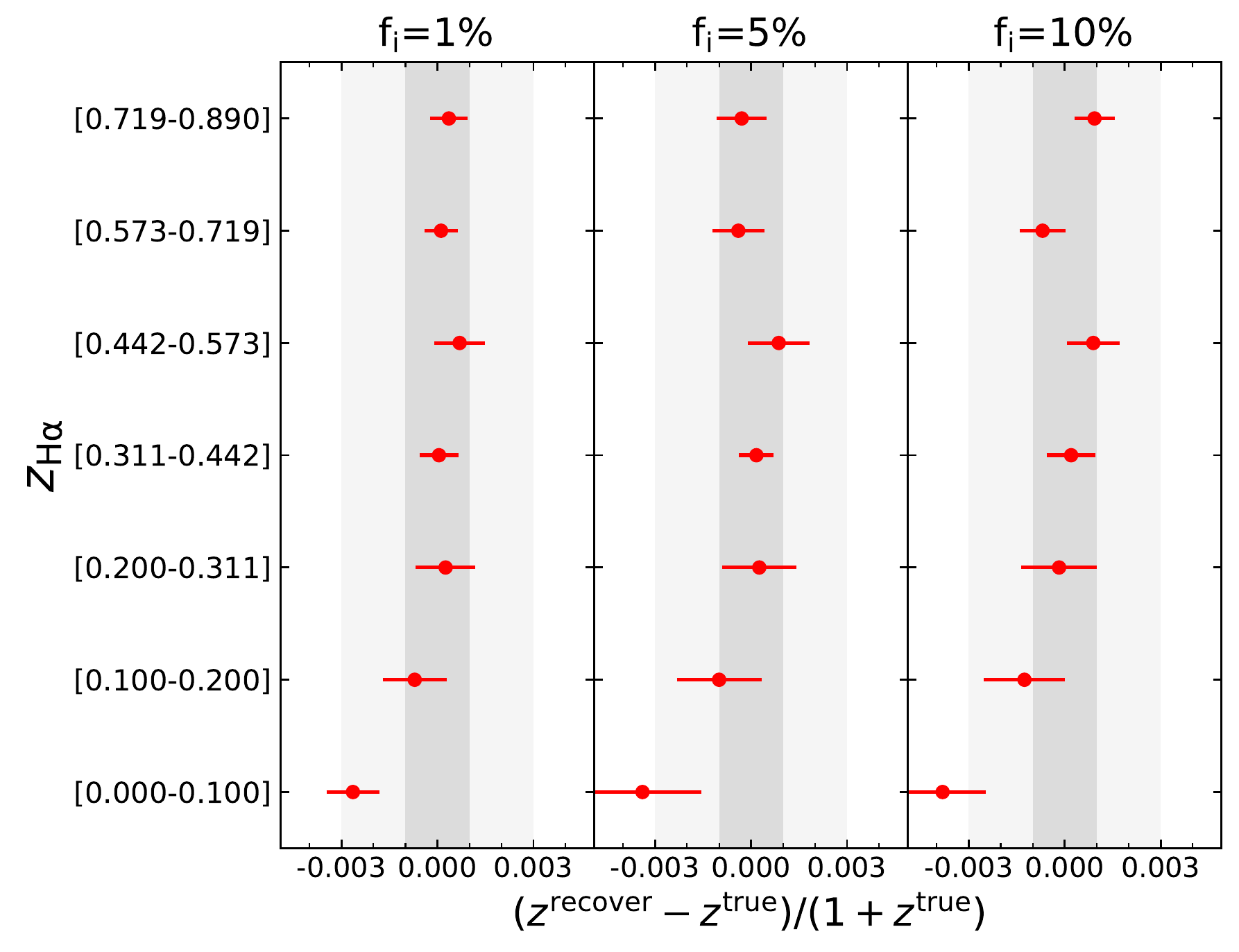}
    \end{minipage}
    % \hspace{1cm}
	\begin{minipage}{8.5cm}
    	\centering
        \includegraphics[width=8.5cm]{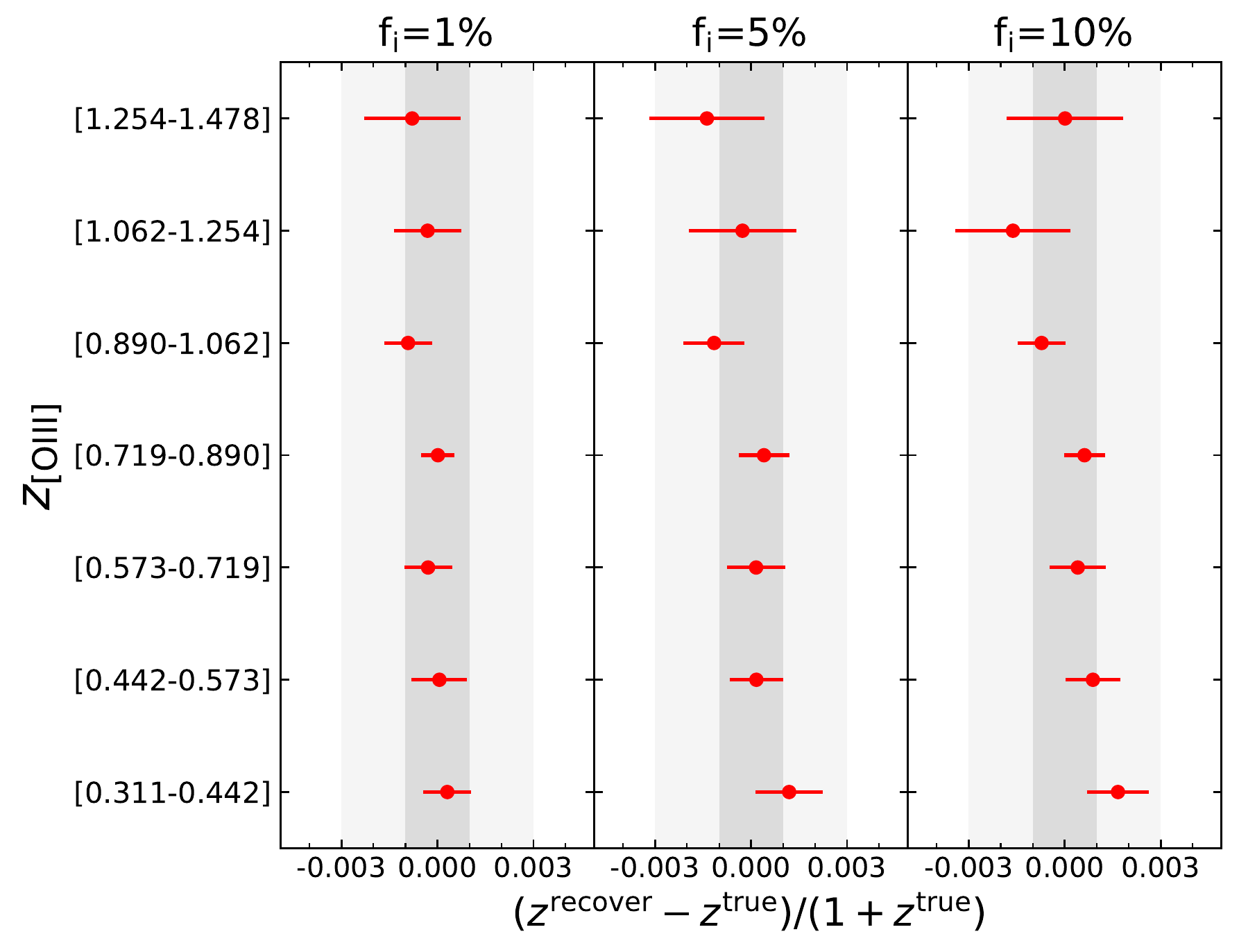}
    \end{minipage}
    \caption{Bias on the mean redshift in each tomographic bin estimated from equation~(\ref{eqn:estimate}) with the reconstructed results when using $\rm H\alpha$ (left panel) and $\rm [O\,\textsc{iii}]$ (right panel) lines to determine the redshifts of galaxies. The points and error bars indicate the bias and $1\sigma$ uncertainties of the estimated mean redshift for each tomographic bin in different cases. The grey and light grey region show the accuracy of $\delta\langle z \rangle < 0.001(1+z)$ and $\delta\langle z \rangle < 0.003(1+z)$, respectively.}
    \label{fig:result_mean_z}
\end{figure*}

In this work, we take the interloper fractions at 1~per cent, 5~per cent and 10~per cent as examples.
We assess the performance of the self-calibration algorithm when using $\rm {H\alpha}$ and $\rm [O\,\textsc{iii}]$ lines to determine the redshifts of galaxies in mock catalogues, respectively.
We still input $\ell C_{\ell}^{gg,D}$ with 1000 initial matrices and the solution selection criterion is also the same as in \citet{Peng:2022aa} via
\begin{equation}
    \frac{\mathcal{J}-\mathcal{J}_{\min}}{\mathcal{J}_{\min}}\leq 10~\text{per cent}\ .
    \label{define:J_selection}
\end{equation}
Here, $\mathcal{J}_{\min}$ is smallest $\mathcal{J}$ from the modified algorithm.
We take the average value of the solutions in this range as the final result.
However, the reconstruction uncertainty here is not decided by the standard deviation of the selected solutions like \citet{Peng:2022aa}.
Since we have much fewer unknowns in this problem, the solutions after selection are very concentrated.
The variance among the selected results is not a good estimate
of the uncertainties induced by the reconstruction algorithm.
Instead, we divide the maps into 10 groups, each representing a sky coverage over $\sim$~1500~$\rm {deg^2}$.
Then we use the standard deviation of the reconstruction results from these 10 groups, divided by $\sqrt{10}$, to represent the uncertainty of the complete sample.
% we scale the error on these measurements by a factor of $\sqrt{10}$ in the following analyses.

As shown in Fig.~\ref{fig:example}, the interlopers only happen in seven observed redshift bins in our mock data,
bin 1-7 and bin 4-10 when the redshifts are determined by $\rm {H\alpha}$ and $\rm [O\,\textsc{iii}]$ lines, respectively.
The points and error bars in Fig.~\ref{fig:result_interloper} show the reconstruction results in different cases when ignoring the cosmic magnification.
Here we only present the reconstructed off-diagonal elements, i.e. interloper fraction $f_{\rm i}$ in each tomographic bin, because of the column-sum-to-one constrain.
We find that the biases of reconstructed interloper fractions are very small except the first redshift bin when using $\rm H\alpha$ to determine the redshift.
This underestimation also occurs when the method in \citet{Gong:2021tb} is used (see Appendix \ref{appendix_A}).
We argue that this deviation is partially due to the low $\rm S/N$ of power spectrum measurement for the lowest redshift bin, and partially due to the cosmic variance. 
It is not a great concern here.
We also find that the reconstruction accuracy is not sensitive to the different interloper fractions we set.
Taking all three cases together, the mean absolute bias of the reconstructed interloper fractions is 0.0017 for the remaining six redshift bins when using $\rm H\alpha$ lines to determine the redshifts and 0.0021 for all seven redshift bins when redshifts are determined by [O\,\textsc{iii}] lines.

Moreover, with the interloper fractions reconstructed by the self-calibration algorithm, the bias in the mean redshift estimation for each tomographic bin can be reduced.
We use the approximation equation proposed by
\citet{Zhang:2010wr} to estimate the mean true redshift for a observed bin $i$, which can be expressed as follows:
\begin{equation}
      \langle z_{i}\rangle \simeq \sum_{j} P_{j i}\langle z_{j}^{D}\rangle\ .
      \label{eqn:estimate}
\end{equation}
Here, $\langle z_{j}^{D}\rangle$ is the mean observed redshift for the observed redshift bin $j$, and it is assumed to be approximately equal to the unknown mean true redshift of the true redshift bin $j$.
The bias of the mean redshift in each tomographic bin obtained by equation~(\ref{eqn:estimate}) are shown in Fig.~\ref{fig:result_mean_z}.
The error bars again come from the standard deviation of the mean redshifts estimated from 10 groups, and divided by $\sqrt{10}$.
We can see that the deviation from the truth value in each tomographic bin has been successfully reduced to $\sim$~0.001$(1+z)$.
% meet the LSST\footnote{LSST, DESI, \textit{RST}, \textit{Euclid}, and \textit{CSST} are classified as Stage IV surveys.} requirements (Y10) of $\delta\langle z \rangle < 0.001(1+z)$ for weak lensing (grey region) and $\delta\langle z \rangle < 0.003(1+z)$ for LSS (light grey region) \citep{The-LSST-Dark-Energy-Science-Collaboration:2018vo}.

\subsection{Cosmic magnification}
\label{sec:magnification}
\begin{figure*}
\centering
    \begin{minipage}{8cm}
        \centering
        \includegraphics[width=8cm]{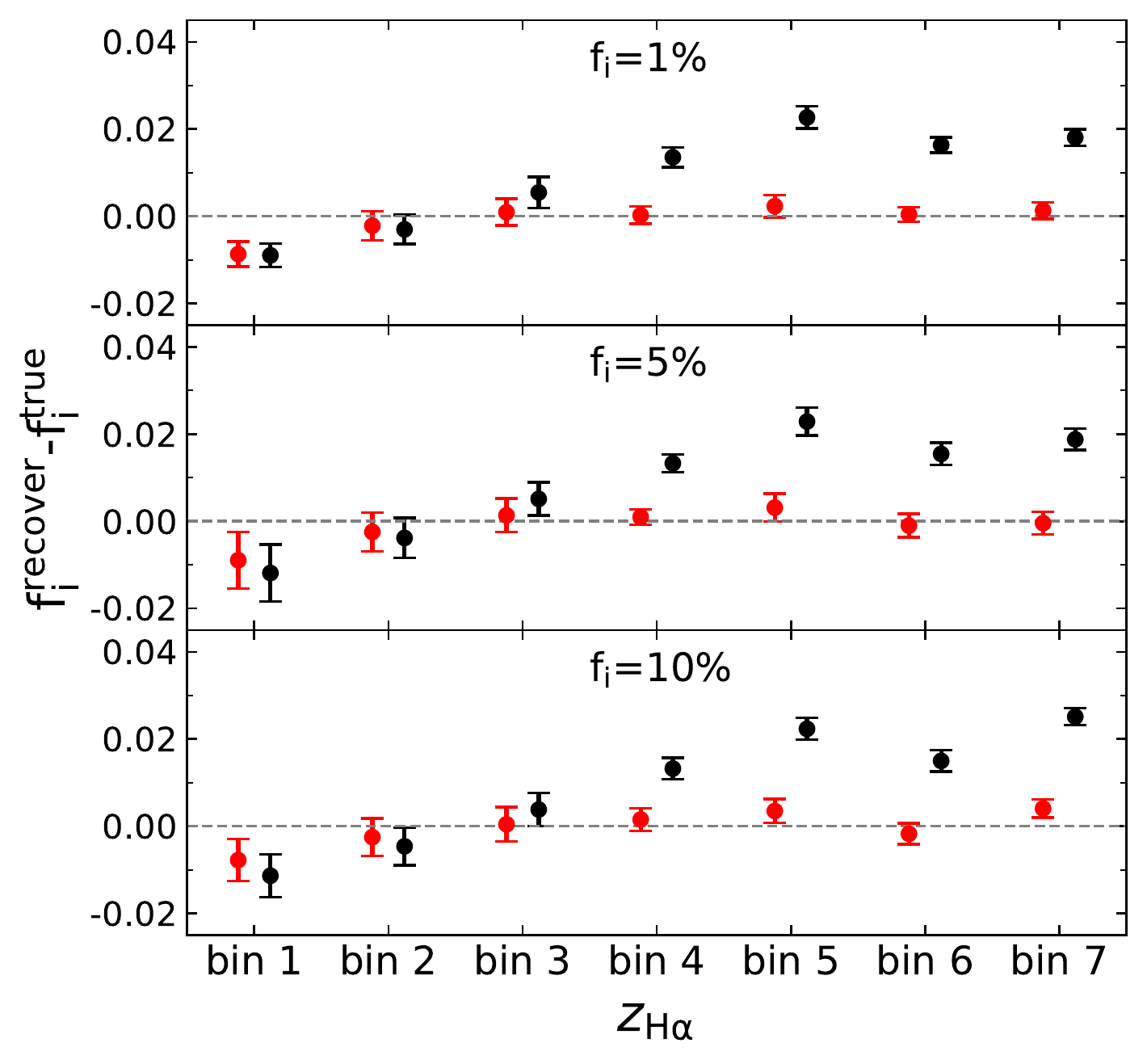}
    \end{minipage}
    \hspace{0.3cm}
	\begin{minipage}{8cm}
    	\centering
        \includegraphics[width=8cm]{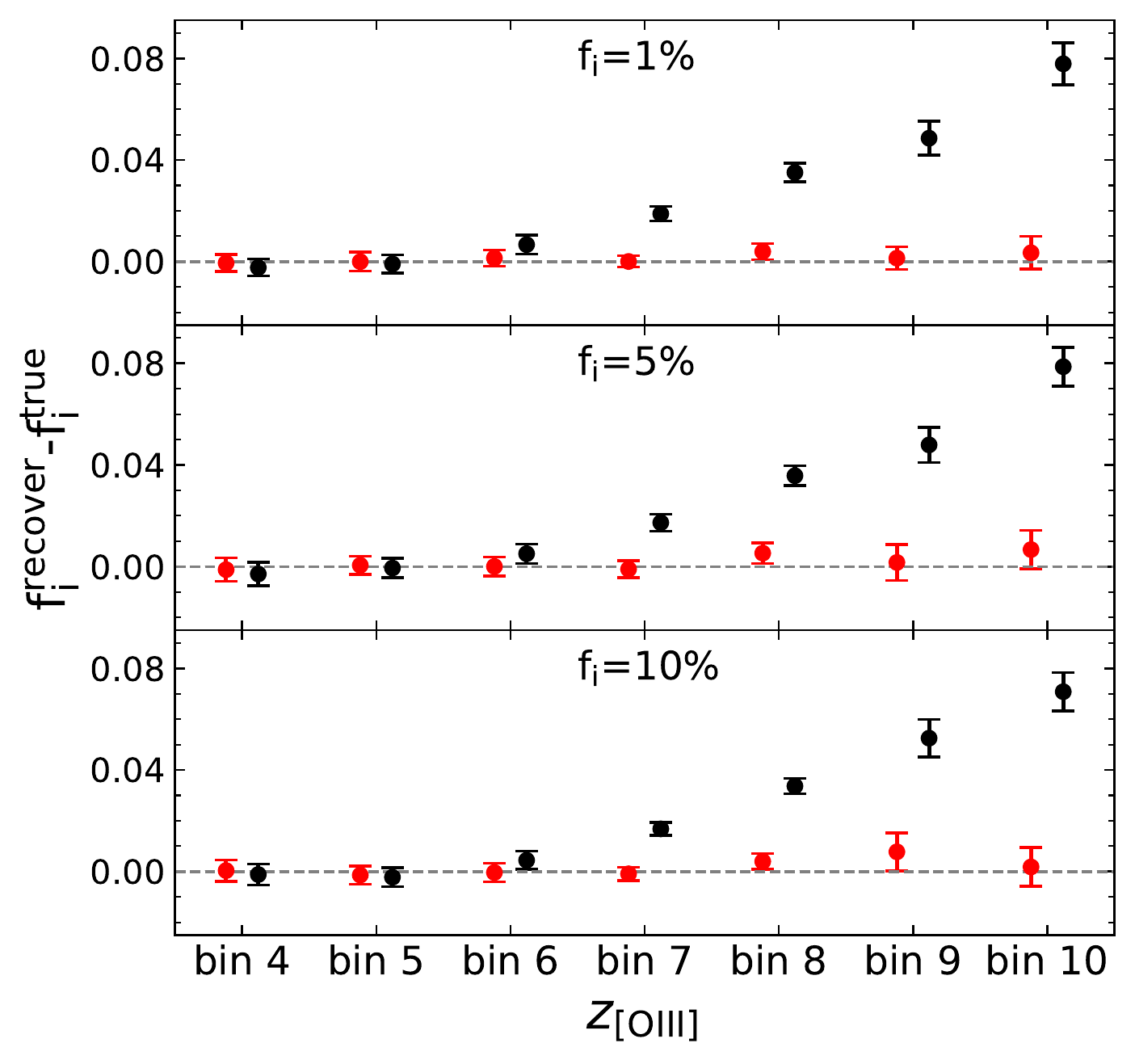}
    \end{minipage}
    \caption{Comparison between the reconstructed interloper fractions after lensing (black points) and the fiducial results before lensing (red points). The plot indicates that the cosmic magnification can drastically degrade the accuracy and needs to be eliminated.}
    \label{fig:result_interloper_mb}
\end{figure*}

\begin{figure*}
\centering
    \begin{minipage}{8cm}
        \centering
        \includegraphics[width=8cm]{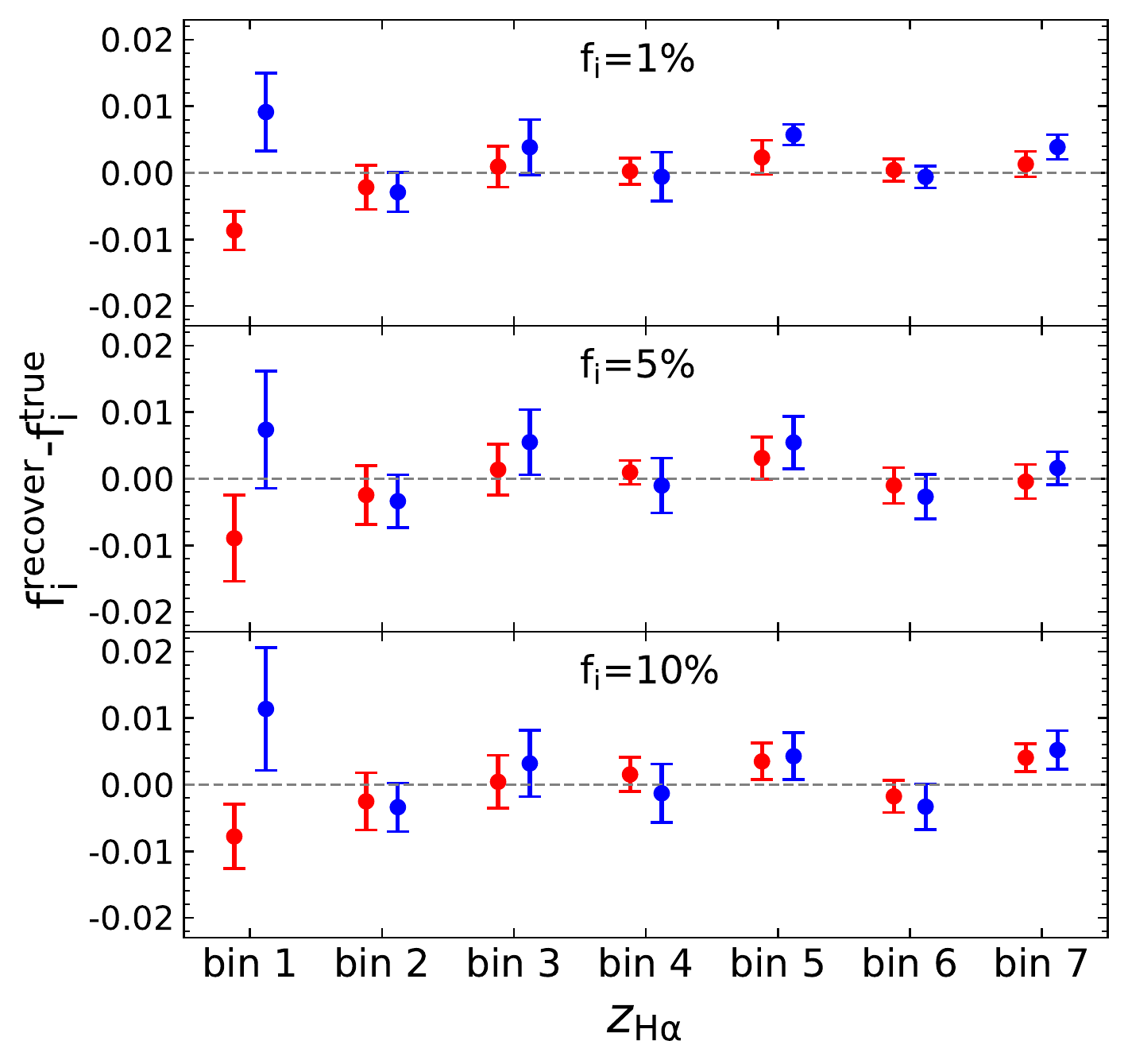}
    \end{minipage}
    \hspace{0.3cm}
	\begin{minipage}{8cm}
    	\centering
        \includegraphics[width=8cm]{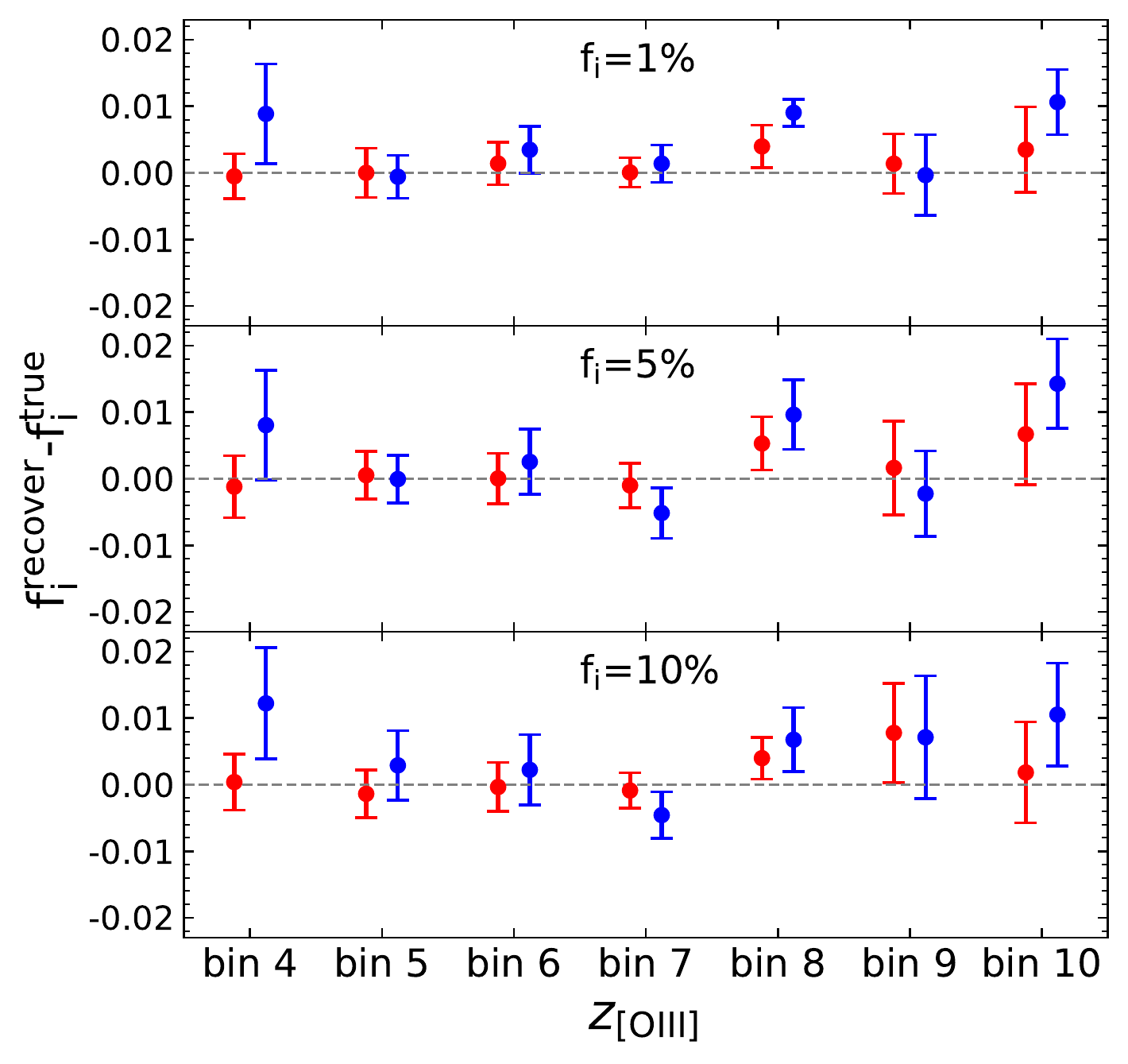}
    \end{minipage}
    \caption{Same as Fig.~\ref{fig:result_interloper_mb}, but the blue points are the results after implementing the elimination method to remove the impact of the cosmic magnification. The comparison shows the effectiveness of our elimination method and the biases of interloper fractions can be reduced to a similar level of the fiducial results before lensing.}
    \label{fig:result_interloper_eliminate}
\end{figure*}
\begin{figure*}
\centering
    \begin{minipage}{8.5cm}
        \centering
        \includegraphics[width=8.5cm]{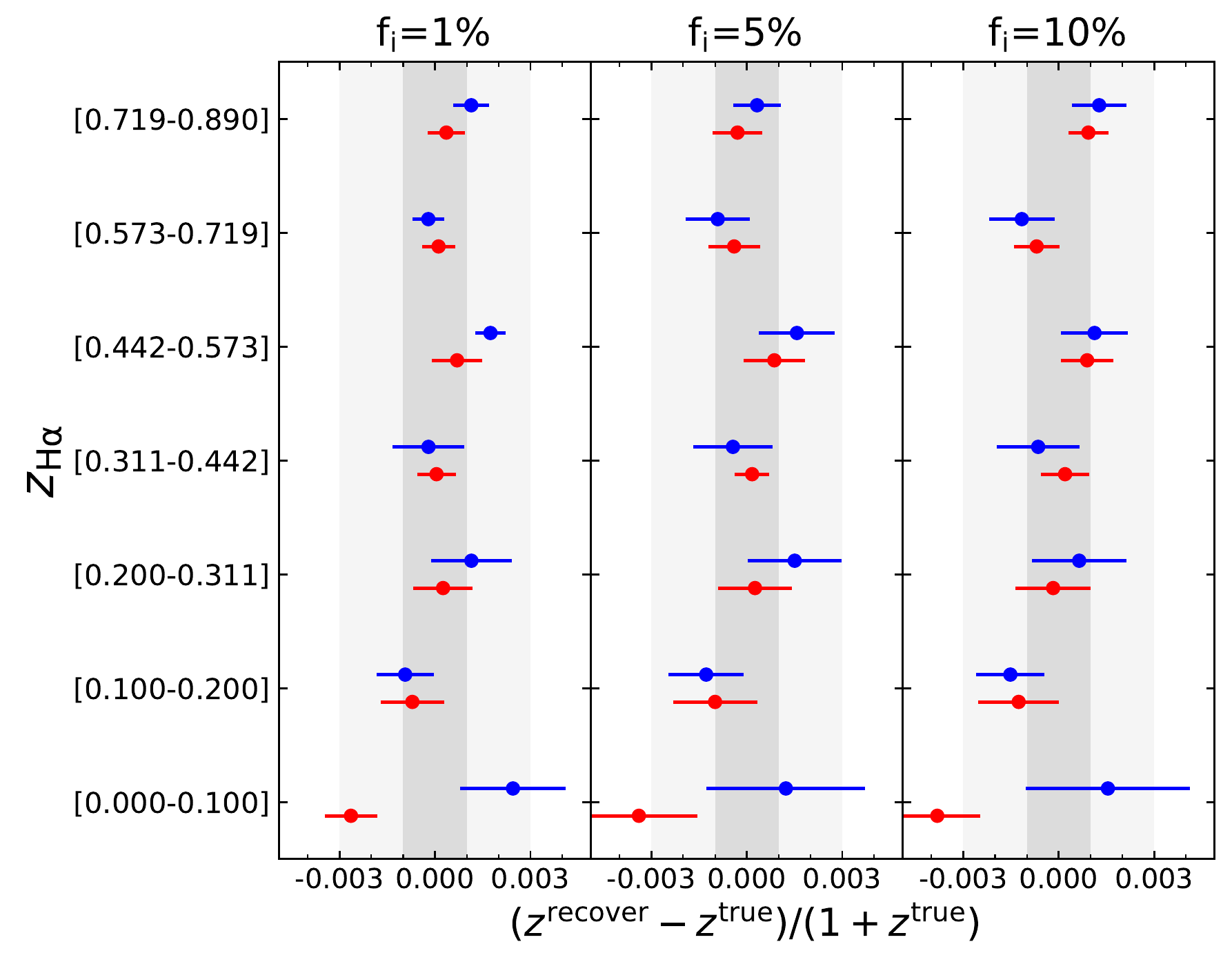}
    \end{minipage}
    % \hspace{1cm}
	\begin{minipage}{8.5cm}
    	\centering
        \includegraphics[width=8.5cm]{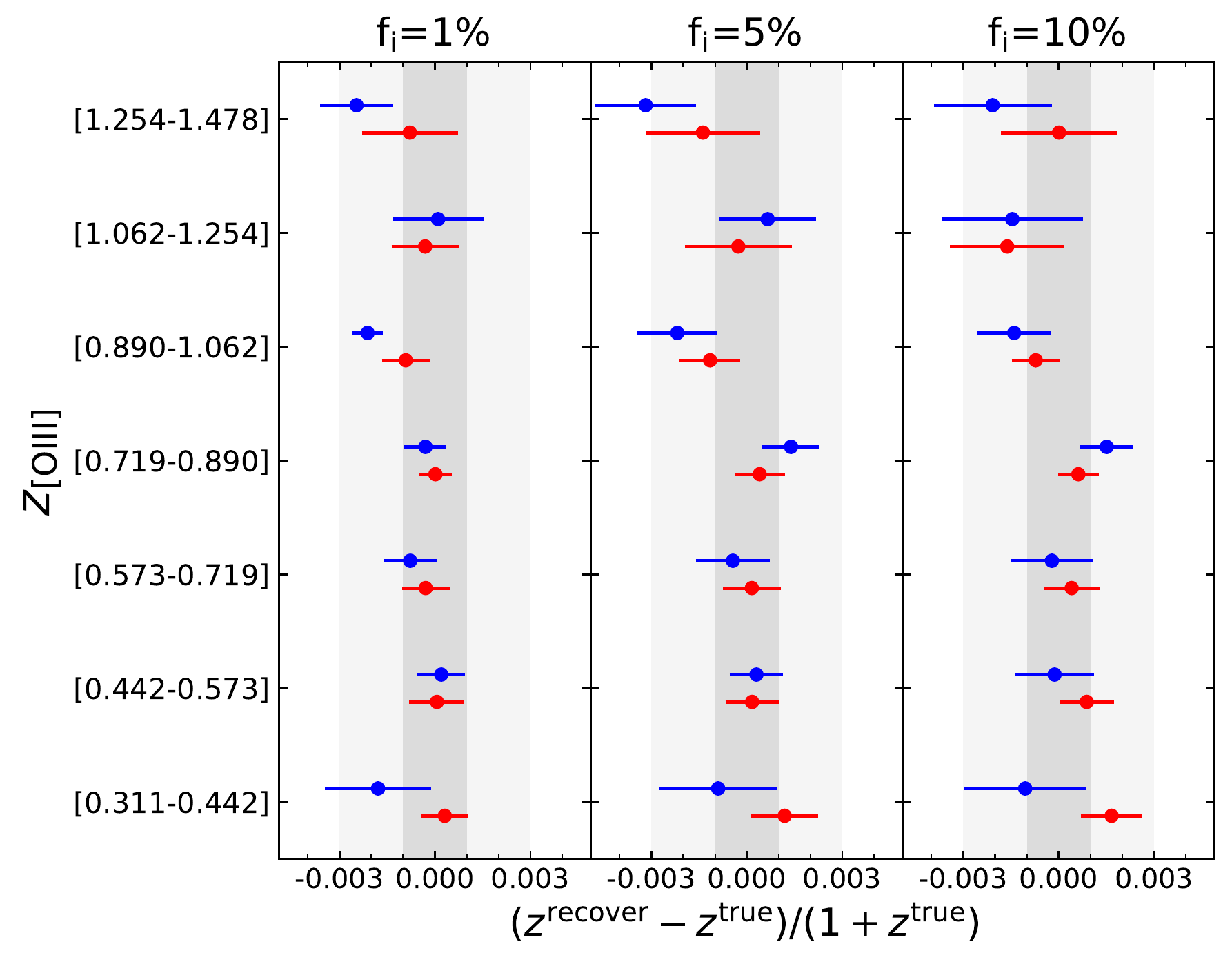}
    \end{minipage}
    \caption{Comparison between the bias on the mean redshift in each tomographic bin after implementing the elimination method to remove the impact of the cosmic magnification (blue points) and the fiducial results before lensing (red points). The results show that the self-calibration accuracy can be effectively recovered with slightly larger uncertainty.}
    \label{fig:result_mean_z_eliminate}
\end{figure*}
In real survey, gravitational lensing can alter the galaxy positions and flux.
Then the observed galaxy clustering will be changed, so called magnification bias, which is an important source of systematic error for the precision cosmology.
The galaxy number overdensity after lensing can be given by
\begin{equation}
    \delta^{g}_{L}=\delta^{g}+g \kappa\ .
    \label{eqn:delta_g}
\end{equation}
Here $\kappa$ is the lensing convergence, with prefactor $g \equiv 2(\alpha-1)$.
$\alpha$ is the logarithmic slope at the flux limit $F_{\text{lim}}$, defined as follows:
\begin{equation}
    \alpha=-\left.\frac{\mathrm{d} \ln n(>F)}{\mathrm{d} \ln F}\right|_{F_{\lim }}\ .
    \label{eqn:alpha}
\end{equation}
With the lensed galaxy distribution, the power spectrum between the observed foreground redshift bin $i$ and background redshift bin $j$ can be approximated by a simple form
\begin{equation}
    C_{ij,L}^{g g, D} \approx C_{ij}^{g g, D}+ g_j C_{ij}^{g \kappa, D}+g_i C_{ji}^{g \kappa, D}\approx C_{ij}^{g g, D}+ g_j C_{ij}^{g \kappa, D}\ ,
    \label{eqn:lensed_cggp}
\end{equation}
where $C_{ij}^{g \kappa, D}$ is the galaxy-galaxy lensing power spectrum between observed redshift bins.
This effect is ignored in past related literature.
However, given the high precision of the reconstructed interloper fractions in the last section, 
we investigate the impact of cosmic magnification in our self-calibration method.

The lensing convergence $\kappa$ in each redshift bin is given by
\begin{equation}
    \kappa=\int \kappa(z_\mathrm{s})n(z_\mathrm{s})dz_\mathrm{s}\ ,
    \label{eqn:kappa}
\end{equation}
where $n(z_\mathrm{s})$ is the distribution of the true galaxy redshifts.
The logarithmic luminosity slope $\alpha$ of ten redshift bins are set to [0.20, 0.26, 0.35, 0.55, 0.90, 1.35, 1.90, 2.35, 2.85, 3.50], which are consistent with \citet{Unruh:2020aa}.
Then we use the galaxy number overdensity after lensing, i.e. equation~(\ref{eqn:delta_g}), to calculate the power spectra and apply the self-calibration algorithm.

In Fig.~\ref{fig:result_interloper_mb}, we compare the reconstructed interloper fractions with and without magnification for all cases. 
It is obvious that the cosmic magnification can drastically degrade the accuracy and need to be particularly taken into account.
The algorithm treats the lensing contribution as the effect of interlopers, and the biases are larger for high redshift bins.
In order to ensure the high precision of the self-calibration method, it's essential to eliminate the magnification effect.

We propose a convenient and efficient elimination method here.
Assume that there exist interloper galaxies between the redshift bins $i$ and $j$.
The $g_j C_{ij}^{g \kappa, D}$ is the lensing magnification term we need to eliminate.
We can easily find two assistant redshift bins, $i_-$ and $i_+$, whose upper and lower boundary is the lower and upper boundary of bin $i$, respectively.
When the widths of redshift bins $i_-$ and $i_+$ are equal to the redshift bin $i$, we have the first order approximation
\begin{equation}
    C_{ij}^{g \kappa, D} \approx \frac{1}{2}(C_{i_-j}^{g \kappa, D}+C_{i_+j}^{g \kappa, D})\ .
    \label{eqn:method_Cgk}
\end{equation}
Note that there are no interlopers between assistant redshift bins $i_\pm$ and redshift bin $j$, thus their cross correlations are  contributed by the galaxy-galaxy lensing only, and we have
\begin{equation}
    g_j C_{ij}^{g \kappa, D} \approx \frac{1}{2}(C_{i_-j, L}^{g g, D}+C_{i_+j, L}^{g g, D})\ .
    \label{eqn:method_Cgg}
\end{equation}
Now we can use the assistant redshift bins to estimate the lensing magnification signal between the redshift bins $i$ and $j$.
And then we can subtract it in the original measured cross power spectrum $C_{ij,L}^{g g, D}$ to obtain $C_{ij}^{g g, D}$.
Note that when the redshift bin $i$ is the first redshift bin, we estimate the magnification signal by working backward from two assistant redshift bins $i_+$ and $i_{++}$ with higher redshift, although this may introduce a relatively large error.
Additionally, in some rare cases involving the first bin, the estimated magnification contribution has the opposite sign to the prefactor $g$ in equation~(\ref{eqn:delta_g}), which is definitely due to the large noise.  We directly ignore the magnification in these situations.
Furthermore, if the bin width is too large to find the equal-width assistant redshift bins, we have to set the width as close as possible and take this difference in bin width into consideration.
% Nevertheless, the above two issues usually only happen for the two lowest redshift bins, where the magnification is least important as shown in Fig.~\ref{fig:result_interloper_mb}

Figs~\ref{fig:result_interloper_eliminate} and \ref{fig:result_mean_z_eliminate} show the results after implementing the elimination method.
In Fig.~\ref{fig:result_interloper_eliminate}, we can see the accuracy of the reconstructed interloper fractions is effectively recovered with slightly larger uncertainty ($\sim$~50~per cent increase).
This indicates that our elimination method can efficiently remove the impact of the cosmic magnification and be applied to ensure a high reconstruction accuracy in practice.
And Fig.~\ref{fig:result_mean_z_eliminate} shows the accuracy of mean redshift in each tomographic bin can also be recovered to a similar level as the fiducial results.
Therefore, although there exist the influence from the cosmic magnification, the reconstruction results with our elimination method can still reach a similar high accuracy.

\section{Conclusions and Discussion}
\label{sec:conclusion}
We modified the algorithm that was applied to self-calibrate the photometric redshift errors, and used it to calibrate the interloper bias in spectroscopic surveys.
The algorithm was validated on the mock data based on \textit{CSST}, with precise reconstructions on the interloper fraction and mean redshift in each tomographic bin.
Actually, the power spectra in true redshift bins can be precisely derived simultaneously.
We also found the cosmic magnification can be a big issue and drastically degrade the accuracy of self-calibration results.
Thus, we proposed a convenient and efficient elimination method to ensure the high accuracy in practice.
After implementing the elimination method, the biases of the reconstructed interloper fractions can be successfully reduced to a similar level as before.
This will play a crucial role in many practical analyses.

In this work, we took the H$\alpha$ and [O\,\textsc{iii}] lines as an example to validate the self-calibration method.
They are widely used as the target lines in the future stage IV spectroscopic surveys (e.g. \textit{RST}, \textit{Euclid}, and \textit{CSST}).
Though with different redshift range, the excellent stability of the reconstruction accuracy across different redshift bins indicate that the self-calibration algorithm and elimination method can also be particularly effective for all these surveys.
And it is obvious that this method is equally applicable to calibrate the interloper contamination between other emission lines, such as the Ly$\alpha$ and [O\,\textsc{ii}] in HETDEX, where approximately 95 per cent of emission line detections are spectra containing only one apparently single-peaked emission line \citep{Davis:2023aa}.
Besides, as long as the corresponding relationship is satisfied, the binning scheme and the range of redshift can be various to match the requirements for different analysis or wavelength coverage from instrument in survey.
We note that our elimination method to remove the magnification impact is not the only approach, however, is the one with minimal assumption.
Any other feasible methods can also be incorporated into the self-calibration process.
For example, a single assistant redshift bin with assuming a cosmology can serve the purpose similarly.

So far our work is carried out based on the simulation data.
The results we obtained give us confidence that our self-calibration algorithm with elimination method presented will enable accurate and well-calibrated redshift for the analyses in \textit{CSST} and other ongoing and future spectroscopic surveys.
We delay the implementation on real data to a future work.

\section*{Acknowledgements}

We thank Pengjie Zhang and Xin Wang for useful discussions.
This work is supported by the National Key Basic Research and Development Program of China (No. 2018YFA0404504), the National Science Foundation of China (grant Nos. 12273020, 11621303, 11890691), the China Manned Space Project with Nos. CMS-CSST-2021-A02 and CMS-CSST-2021-A03, the “111” Project of the Ministry of Education under grant No. B20019, and the sponsorship from Yangyang Development Fund.
This work made use of the Gravity Supercomputer at the Department of Astronomy, Shanghai Jiao Tong University.

\section*{Data availability}
All data included in this study are available upon reasonable request by contacting with the corresponding author.



\bibliographystyle{mnras}
\bibliography{example} % if your bibtex file is called example.bib

\begin{thebibliography}{}
\makeatletter
\relax
\def\mn@urlcharsother{\let\do\@makeother \do\$\do\&\do\#\do\^\do\_\do\%\do\~}
\def\mn@doi{\begingroup\mn@urlcharsother \@ifnextchar [ {\mn@doi@}
  {\mn@doi@[]}}
\def\mn@doi@[#1]#2{\def\@tempa{#1}\ifx\@tempa\@empty \href
  {http://dx.doi.org/#2} {doi:#2}\else \href {http://dx.doi.org/#2} {#1}\fi
  \endgroup}
\def\mn@eprint#1#2{\mn@eprint@#1:#2::\@nil}
\def\mn@eprint@arXiv#1{\href {http://arxiv.org/abs/#1} {{\tt arXiv:#1}}}
\def\mn@eprint@dblp#1{\href {http://dblp.uni-trier.de/rec/bibtex/#1.xml}
  {dblp:#1}}
\def\mn@eprint@#1:#2:#3:#4\@nil{\def\@tempa {#1}\def\@tempb {#2}\def\@tempc
  {#3}\ifx \@tempc \@empty \let \@tempc \@tempb \let \@tempb \@tempa \fi \ifx
  \@tempb \@empty \def\@tempb {arXiv}\fi \@ifundefined
  {mn@eprint@\@tempb}{\@tempb:\@tempc}{\expandafter \expandafter \csname
  mn@eprint@\@tempb\endcsname \expandafter{\@tempc}}}

\bibitem[\protect\citeauthoryear{{Addison}, {Bennett}, {Jeong}, {Komatsu}  \&
  {Weiland}}{{Addison} et~al.}{2019}]{Addison:2019aa}
{Addison} G.~E.,  {Bennett} C.~L.,  {Jeong} D.,  {Komatsu} E.,   {Weiland}
  J.~L.,  2019, \mn@doi [\apj] {10.3847/1538-4357/ab22a0}, \href
  {https://ui.adsabs.harvard.edu/abs/2019ApJ...879...15A} {879, 15}

\bibitem[\protect\citeauthoryear{{Alonso}, {Sanchez}, {Slosar}  \& {LSST Dark
  Energy Science Collaboration}}{{Alonso} et~al.}{2019}]{Alonso:2019aa}
{Alonso} D.,  {Sanchez} J.,  {Slosar} A.,   {LSST Dark Energy Science
  Collaboration} 2019, \mn@doi [\mnras] {10.1093/mnras/stz093}, \href
  {https://ui.adsabs.harvard.edu/abs/2019MNRAS.484.4127A} {484, 4127}

\bibitem[\protect\citeauthoryear{{Amendola} et~al.,}{{Amendola}
  et~al.}{2018}]{Amendola:2018aa}
{Amendola} L.,  et~al., 2018, \mn@doi [Living Reviews in Relativity]
  {10.1007/s41114-017-0010-3}, \href
  {https://ui.adsabs.harvard.edu/abs/2018LRR....21....2A} {21, 2}

\bibitem[\protect\citeauthoryear{{Awan} \& {Gawiser}}{{Awan} \&
  {Gawiser}}{2020}]{Awan:2020aa}
{Awan} H.,  {Gawiser} E.,  2020, \mn@doi [\apj] {10.3847/1538-4357/ab63c8},
  \href {https://ui.adsabs.harvard.edu/abs/2020ApJ...890...78A} {890, 78}

\bibitem[\protect\citeauthoryear{{Benjamin}, {van Waerbeke}, {M{\'e}nard}  \&
  {Kilbinger}}{{Benjamin} et~al.}{2010}]{Benjamin:2010aa}
{Benjamin} J.,  {van Waerbeke} L.,  {M{\'e}nard} B.,   {Kilbinger} M.,  2010,
  \mn@doi [\mnras] {10.1111/j.1365-2966.2010.17191.x}, \href
  {https://ui.adsabs.harvard.edu/abs/2010MNRAS.408.1168B} {408, 1168}

\bibitem[\protect\citeauthoryear{{DESI Collaboration} et~al.,}{{DESI
  Collaboration} et~al.}{2016a}]{DESI-Collaboration:2016vy}
{DESI Collaboration} et~al., 2016a, arXiv e-prints, \href
  {https://ui.adsabs.harvard.edu/abs/2016arXiv161100036D} {p. arXiv:1611.00036}

\bibitem[\protect\citeauthoryear{{DESI Collaboration} et~al.,}{{DESI
  Collaboration} et~al.}{2016b}]{DESI-Collaboration:2016vs}
{DESI Collaboration} et~al., 2016b, arXiv e-prints, \href
  {https://ui.adsabs.harvard.edu/abs/2016arXiv161100037D} {p. arXiv:1611.00037}

\bibitem[\protect\citeauthoryear{{Davis} et~al.,}{{Davis}
  et~al.}{2023}]{Davis:2023aa}
{Davis} D.,  et~al., 2023, \mn@doi [\apj] {10.3847/1538-4357/acb0ca}, \href
  {https://ui.adsabs.harvard.edu/abs/2023ApJ...946...86D} {946, 86}

\bibitem[\protect\citeauthoryear{{Farrow} et~al.,}{{Farrow}
  et~al.}{2021}]{Farrow:2021aa}
{Farrow} D.~J.,  et~al., 2021, \mn@doi [\mnras] {10.1093/mnras/stab1986}, \href
  {https://ui.adsabs.harvard.edu/abs/2021MNRAS.507.3187F} {507, 3187}

\bibitem[\protect\citeauthoryear{{Foroozan}, {Massara}  \&
  {Percival}}{{Foroozan} et~al.}{2022}]{Foroozan:2022aa}
{Foroozan} S.,  {Massara} E.,   {Percival} W.~J.,  2022, \mn@doi [\jcap]
  {10.1088/1475-7516/2022/10/072}, \href
  {https://ui.adsabs.harvard.edu/abs/2022JCAP...10..072F} {2022, 072}

\bibitem[\protect\citeauthoryear{{Gebhardt} et~al.,}{{Gebhardt}
  et~al.}{2021}]{Gebhardt:2021aa}
{Gebhardt} K.,  et~al., 2021, \mn@doi [\apj] {10.3847/1538-4357/ac2e03}, \href
  {https://ui.adsabs.harvard.edu/abs/2021ApJ...923..217G} {923, 217}

\bibitem[\protect\citeauthoryear{{Gong} et~al.,}{{Gong}
  et~al.}{2019}]{Gong:2019tb}
{Gong} Y.,  et~al., 2019, \mn@doi [\apj] {10.3847/1538-4357/ab391e}, \href
  {https://ui.adsabs.harvard.edu/abs/2019ApJ...883..203G} {883, 203}

\bibitem[\protect\citeauthoryear{{Gong}, {Miao}, {Zhang}  \& {Chen}}{{Gong}
  et~al.}{2021}]{Gong:2021tb}
{Gong} Y.,  {Miao} H.,  {Zhang} P.,   {Chen} X.,  2021, \mn@doi [\apj]
  {10.3847/1538-4357/ac1350}, \href
  {https://ui.adsabs.harvard.edu/abs/2021ApJ...919...12G} {919, 12}

\bibitem[\protect\citeauthoryear{{G{\'o}rski}, {Hivon}, {Banday}, {Wandelt},
  {Hansen}, {Reinecke}  \& {Bartelmann}}{{G{\'o}rski}
  et~al.}{2005}]{Gorski:2005te}
{G{\'o}rski} K.~M.,  {Hivon} E.,  {Banday} A.~J.,  {Wandelt} B.~D.,  {Hansen}
  F.~K.,  {Reinecke} M.,   {Bartelmann} M.,  2005, \mn@doi [\apj]
  {10.1086/427976}, \href
  {https://ui.adsabs.harvard.edu/abs/2005ApJ...622..759G} {622, 759}

\bibitem[\protect\citeauthoryear{{Grasshorn Gebhardt} et~al.,}{{Grasshorn
  Gebhardt} et~al.}{2019}]{Grasshorn-Gebhardt:2019aa}
{Grasshorn Gebhardt} H.~S.,  et~al., 2019, \mn@doi [\apj]
  {10.3847/1538-4357/ab12d5}, \href
  {https://ui.adsabs.harvard.edu/abs/2019ApJ...876...32G} {876, 32}

\bibitem[\protect\citeauthoryear{{Hinshaw} et~al.,}{{Hinshaw}
  et~al.}{2013}]{Hinshaw:2013vg}
{Hinshaw} G.,  et~al., 2013, \mn@doi [\apjs] {10.1088/0067-0049/208/2/19},
  \href {https://ui.adsabs.harvard.edu/abs/2013ApJS..208...19H} {208, 19}

\bibitem[\protect\citeauthoryear{{Jing}}{{Jing}}{2019}]{Jing:2019uw}
{Jing} Y.,  2019, \mn@doi [Science China Physics, Mechanics, and Astronomy]
  {10.1007/s11433-018-9286-x}, \href
  {https://ui.adsabs.harvard.edu/abs/2019SCPMA..6219511J} {62, 19511}

\bibitem[\protect\citeauthoryear{{Kirby}, {Guhathakurta}, {Faber}, {Koo},
  {Weiner}  \& {Cooper}}{{Kirby} et~al.}{2007}]{Kirby:2007aa}
{Kirby} E.~N.,  {Guhathakurta} P.,  {Faber} S.~M.,  {Koo} D.~C.,  {Weiner}
  B.~J.,   {Cooper} M.~C.,  2007, \mn@doi [\apj] {10.1086/513464}, \href
  {https://ui.adsabs.harvard.edu/abs/2007ApJ...660...62K} {660, 62}

\bibitem[\protect\citeauthoryear{{Komatsu} et~al.,}{{Komatsu}
  et~al.}{2011}]{Komatsu:2011un}
{Komatsu} E.,  et~al., 2011, \mn@doi [\apjs] {10.1088/0067-0049/192/2/18},
  \href {https://ui.adsabs.harvard.edu/abs/2011ApJS..192...18K} {192, 18}

\bibitem[\protect\citeauthoryear{{Leung} et~al.,}{{Leung}
  et~al.}{2017}]{Leung:2017aa}
{Leung} A.~S.,  et~al., 2017, \mn@doi [\apj] {10.3847/1538-4357/aa71af}, \href
  {https://ui.adsabs.harvard.edu/abs/2017ApJ...843..130L} {843, 130}

\bibitem[\protect\citeauthoryear{{Lilly} et~al.,}{{Lilly}
  et~al.}{2007}]{Lilly:2007wd}
{Lilly} S.~J.,  et~al., 2007, \mn@doi [\apjs] {10.1086/516589}, \href
  {https://ui.adsabs.harvard.edu/abs/2007ApJS..172...70L} {172, 70}

\bibitem[\protect\citeauthoryear{{Lilly} et~al.,}{{Lilly}
  et~al.}{2009}]{Lilly:2009vb}
{Lilly} S.~J.,  et~al., 2009, \mn@doi [\apjs] {10.1088/0067-0049/184/2/218},
  \href {https://ui.adsabs.harvard.edu/abs/2009ApJS..184..218L} {184, 218}

\bibitem[\protect\citeauthoryear{{Massara}, {Ho}, {Hirata}, {DeRose},
  {Wechsler}  \& {Fang}}{{Massara} et~al.}{2021}]{Massara:2021aa}
{Massara} E.,  {Ho} S.,  {Hirata} C.~M.,  {DeRose} J.,  {Wechsler} R.~H.,
  {Fang} X.,  2021, \mn@doi [\mnras] {10.1093/mnras/stab2628}, \href
  {https://ui.adsabs.harvard.edu/abs/2021MNRAS.508.4193M} {508, 4193}

\bibitem[\protect\citeauthoryear{{Mentuch Cooper} et~al.,}{{Mentuch Cooper}
  et~al.}{2023}]{Mentuch-Cooper:2023aa}
{Mentuch Cooper} E.,  et~al., 2023, \mn@doi [\apj] {10.3847/1538-4357/aca962},
  \href {https://ui.adsabs.harvard.edu/abs/2023ApJ...943..177M} {943, 177}

\bibitem[\protect\citeauthoryear{{Peng}, {Xu}, {Zhang}, {Chen}  \& {Yu}}{{Peng}
  et~al.}{2022}]{Peng:2022aa}
{Peng} H.,  {Xu} H.,  {Zhang} L.,  {Chen} Z.,   {Yu} Y.,  2022, \mn@doi
  [\mnras] {10.1093/mnras/stac2713}, \href
  {https://ui.adsabs.harvard.edu/abs/2022MNRAS.516.6210P} {516, 6210}

\bibitem[\protect\citeauthoryear{{Pullen}, {Hirata}, {Dor{\'e}}  \&
  {Raccanelli}}{{Pullen} et~al.}{2016}]{Pullen:2016vt}
{Pullen} A.~R.,  {Hirata} C.~M.,  {Dor{\'e}} O.,   {Raccanelli} A.,  2016,
  \mn@doi [\pasj] {10.1093/pasj/psv118}, \href
  {https://ui.adsabs.harvard.edu/abs/2016PASJ...68...12P} {68, 12}

\bibitem[\protect\citeauthoryear{{Schaan}, {Ferraro}  \& {Seljak}}{{Schaan}
  et~al.}{2020}]{Schaan:2020up}
{Schaan} E.,  {Ferraro} S.,   {Seljak} U.,  2020, \mn@doi [\jcap]
  {10.1088/1475-7516/2020/12/001}, \href
  {https://ui.adsabs.harvard.edu/abs/2020JCAP...12..001S} {2020, 001}

\bibitem[\protect\citeauthoryear{{Schneider}, {Knox}, {Zhan}  \&
  {Connolly}}{{Schneider} et~al.}{2006}]{Schneider:2006ta}
{Schneider} M.,  {Knox} L.,  {Zhan} H.,   {Connolly} A.,  2006, \mn@doi [\apj]
  {10.1086/507675}, \href
  {https://ui.adsabs.harvard.edu/abs/2006ApJ...651...14S} {651, 14}

\bibitem[\protect\citeauthoryear{{Spergel} et~al.,}{{Spergel}
  et~al.}{2015}]{Spergel:2015aa}
{Spergel} D.,  et~al., 2015, arXiv e-prints, \href
  {https://ui.adsabs.harvard.edu/abs/2015arXiv150303757S} {p. arXiv:1503.03757}

\bibitem[\protect\citeauthoryear{{Sun}, {Zhang}, {Dong}, {Yao}, {Shan},
  {Jullo}, {Kneib}  \& {Yin}}{{Sun} et~al.}{2023}]{Sun:2023aa}
{Sun} Z.,  {Zhang} P.,  {Dong} F.,  {Yao} J.,  {Shan} H.,  {Jullo} E.,  {Kneib}
  J.-P.,   {Yin} B.,  2023, \mn@doi [\apjs] {10.3847/1538-4365/acda2a}, \href
  {https://ui.adsabs.harvard.edu/abs/2023ApJS..267...21S} {267, 21}

\bibitem[\protect\citeauthoryear{{Takada} et~al.,}{{Takada}
  et~al.}{2014}]{Takada:2014vd}
{Takada} M.,  et~al., 2014, \mn@doi [\pasj] {10.1093/pasj/pst019}, \href
  {https://ui.adsabs.harvard.edu/abs/2014PASJ...66R...1T} {66, R1}

\bibitem[\protect\citeauthoryear{{Unruh}, {Schneider}, {Hilbert}, {Simon},
  {Martin}  \& {Puertas}}{{Unruh} et~al.}{2020}]{Unruh:2020aa}
{Unruh} S.,  {Schneider} P.,  {Hilbert} S.,  {Simon} P.,  {Martin} S.,
  {Puertas} J.~C.,  2020, \mn@doi [\aap] {10.1051/0004-6361/201936915}, \href
  {https://ui.adsabs.harvard.edu/abs/2020A&A...638A..96U} {638, A96}

\bibitem[\protect\citeauthoryear{{Xu} et~al.,}{{Xu} et~al.}{2023}]{Xu:2023aa}
{Xu} H.,  et~al., 2023, \mn@doi [\mnras] {10.1093/mnras/stad136}, \href
  {https://ui.adsabs.harvard.edu/abs/2023MNRAS.520..161X} {520, 161}

\bibitem[\protect\citeauthoryear{{Zhang}, {Pen}  \& {Bernstein}}{{Zhang}
  et~al.}{2010}]{Zhang:2010wr}
{Zhang} P.,  {Pen} U.-L.,   {Bernstein} G.,  2010, \mn@doi [\mnras]
  {10.1111/j.1365-2966.2010.16445.x}, \href
  {https://ui.adsabs.harvard.edu/abs/2010MNRAS.405..359Z} {405, 359}

\bibitem[\protect\citeauthoryear{{Zhang}, {Yu}  \& {Zhang}}{{Zhang}
  et~al.}{2017}]{Zhang:2017um}
{Zhang} L.,  {Yu} Y.,   {Zhang} P.,  2017, \mn@doi [\apj]
  {10.3847/1538-4357/aa8c72}, \href
  {https://ui.adsabs.harvard.edu/abs/2017ApJ...848...44Z} {848, 44}

\makeatother
\end{thebibliography}

%%%%%%%%%%%%%%%%% APPENDICES %%%%%%%%%%%%%%%%%%%%%

\appendix
\section{Previous method and the improvement from an iterative way}
\label{appendix_A}
\begin{figure*}
\centering
    \begin{minipage}{8.5cm}
        \centering
        \includegraphics[width=8.5cm]{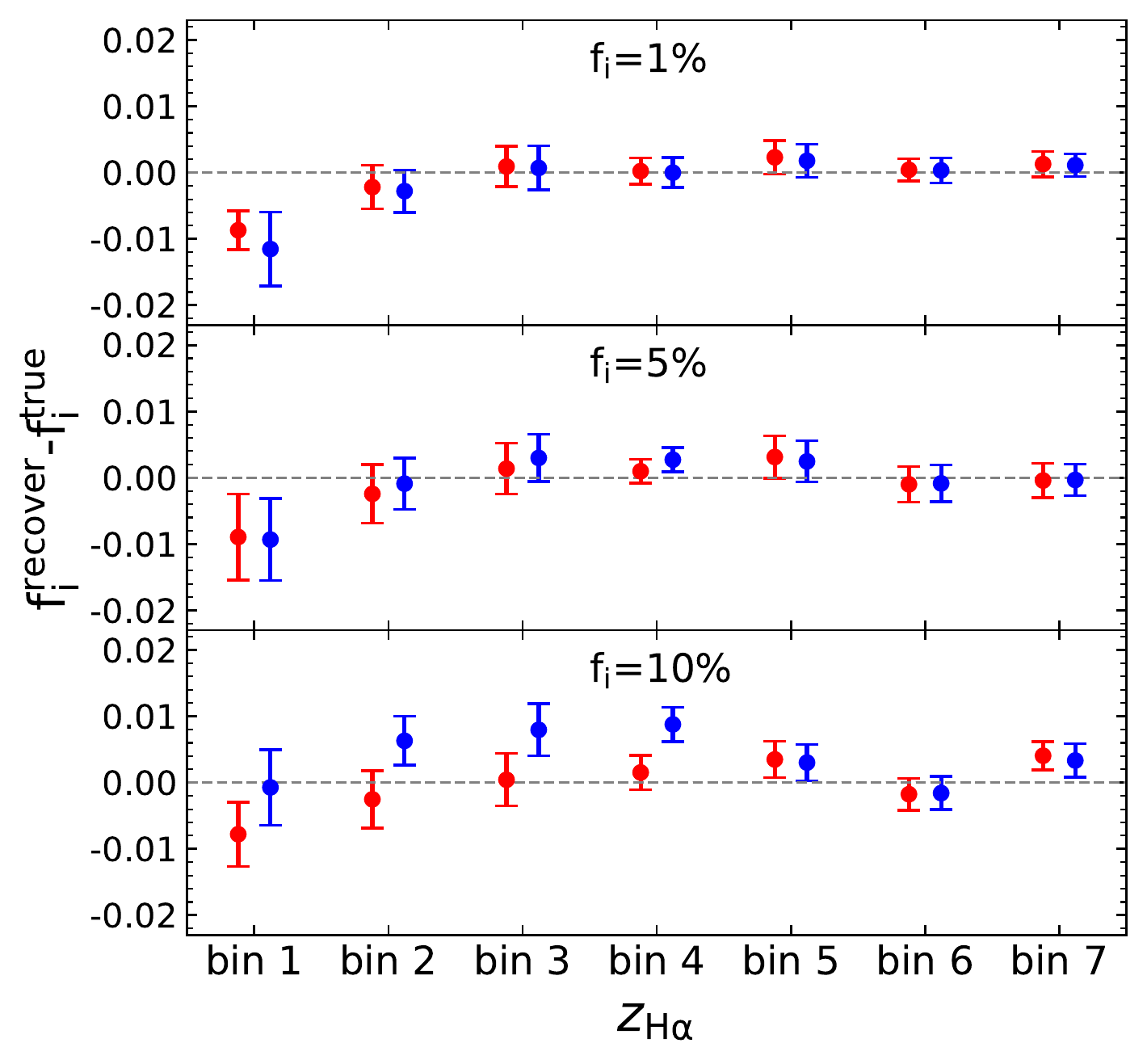}
    \end{minipage}
    \hspace{0.3cm}
	\begin{minipage}{8.5cm}
    	\centering
        \includegraphics[width=8.5cm]{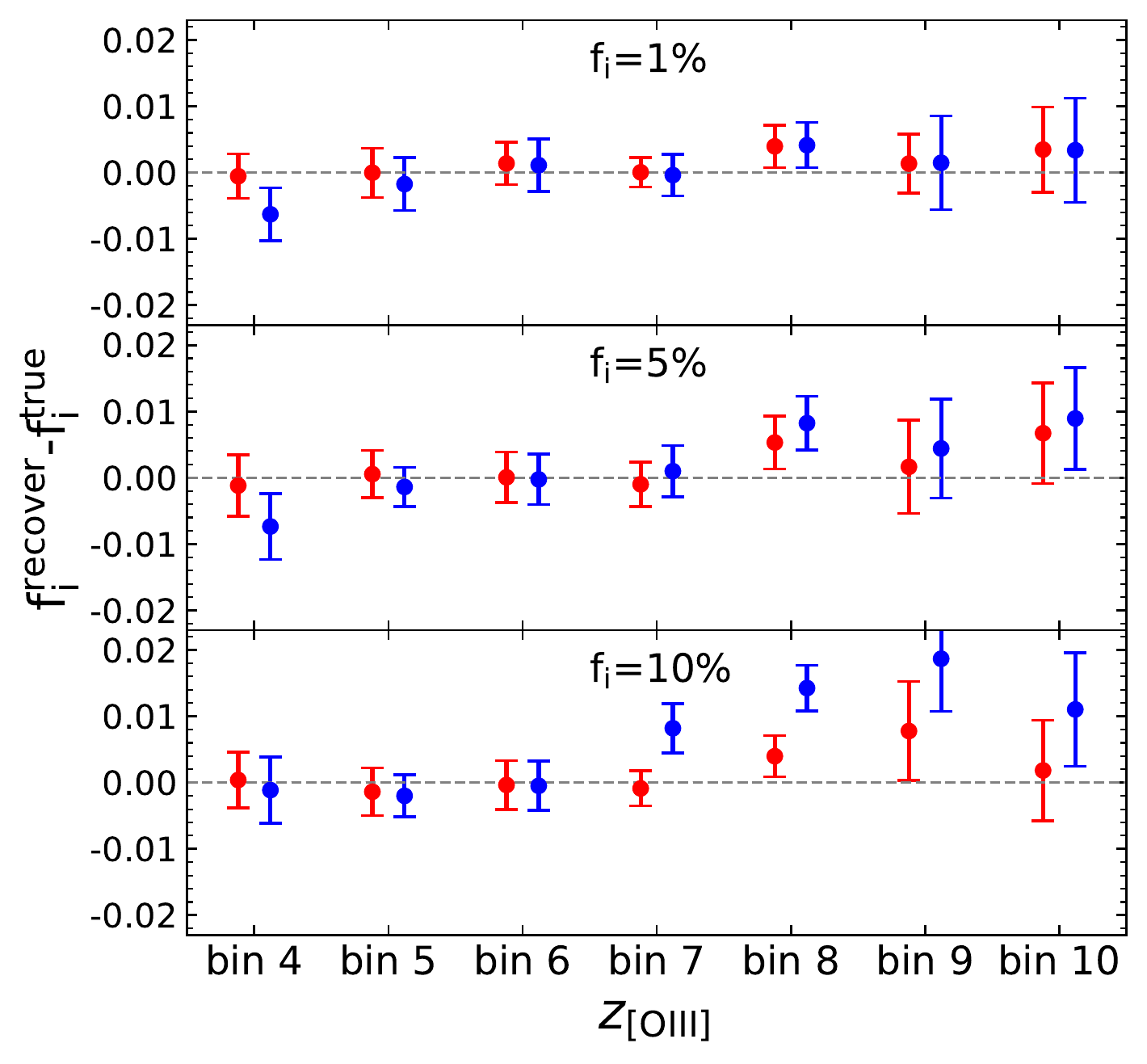}
    \end{minipage}
    \caption{Comparison of results from the self-calibration algorithm (red points) and the ratio method in \citet{Gong:2021tb} (blue points). The plot shows that the reconstruction accuracy in some redshift bins using the previous ratio method will degrade as the interloper fractions rise.}
    \label{fig:result_compare_gongyan}
\end{figure*}

\begin{figure*}
\centering
    \begin{minipage}{8.5cm}
        \centering
        \includegraphics[width=8.5cm]{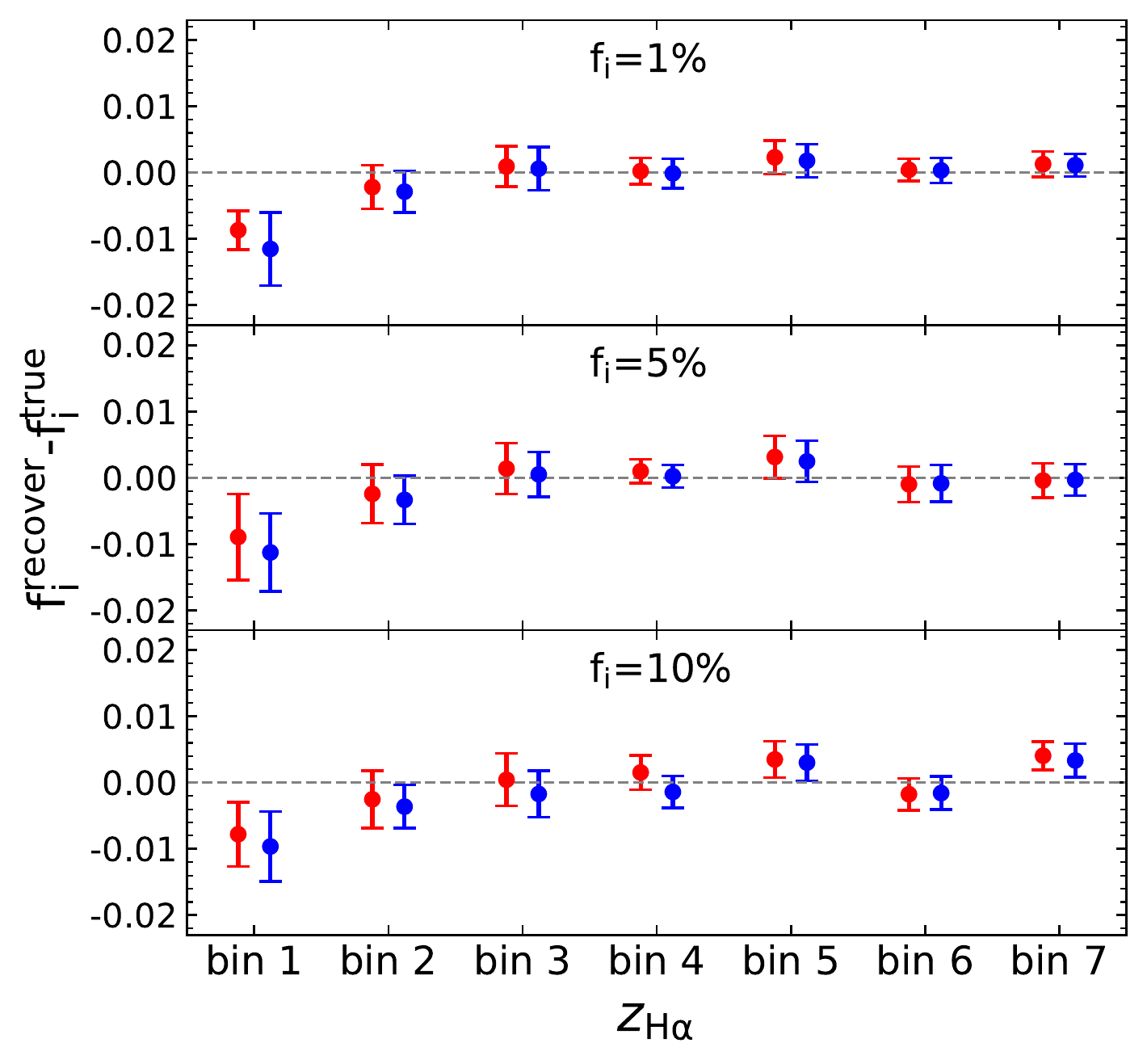}
    \end{minipage}
    \hspace{0.3cm}
	\begin{minipage}{8.5cm}
    	\centering
        \includegraphics[width=8.5cm]{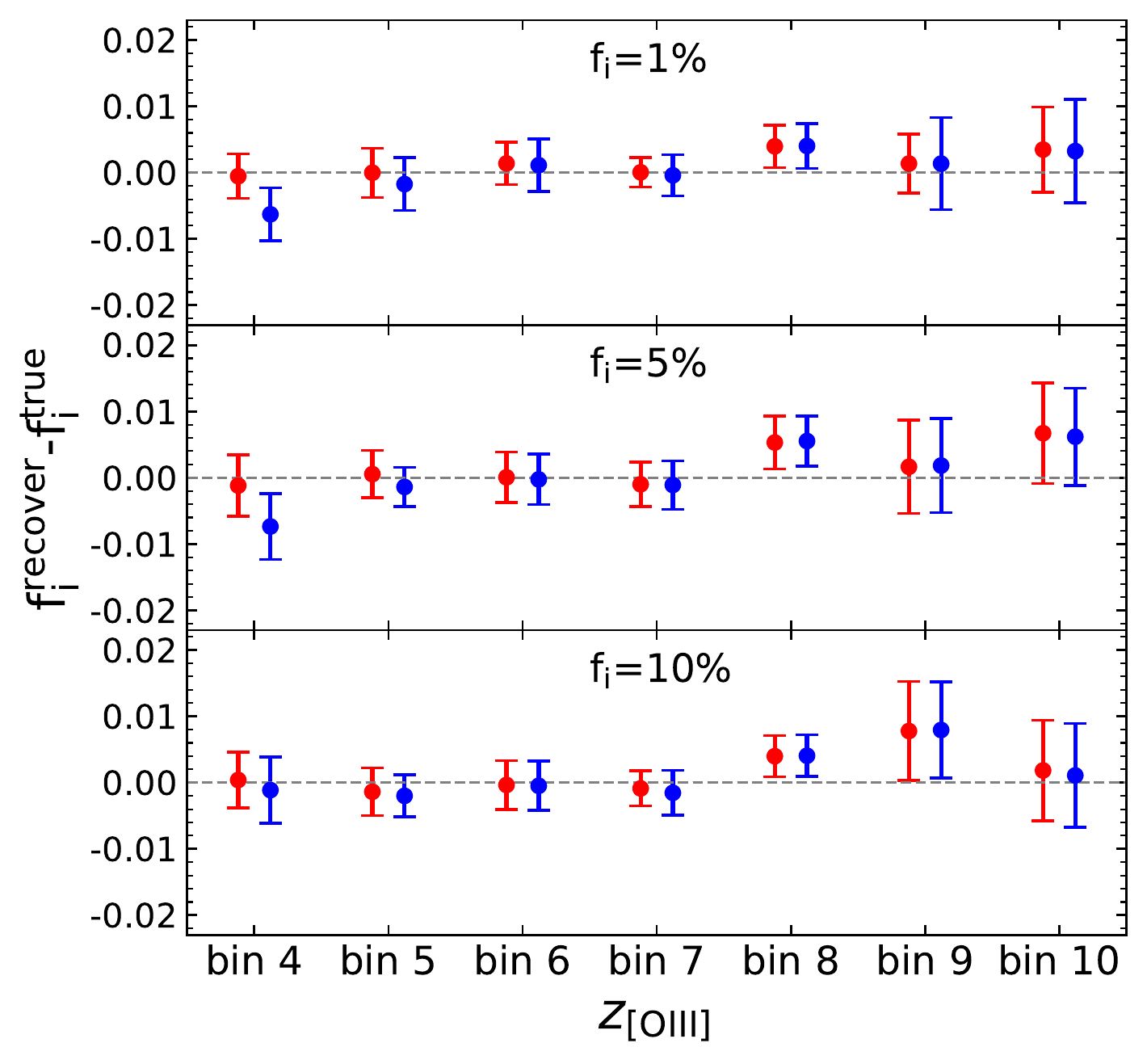}
    \end{minipage}
    \caption{Same as Fig.~\ref{fig:result_compare_gongyan}, but the blue points are further calibrated using the iteration method. The iteration process follows the rule in equation~(\ref{eqn:fraction bias simplified}). The results show that the iteration method can successfully solve the problem in previous ratio method and is comparable to the self-calibration algorithm.}
    \label{fig:result_compare_method}
\end{figure*}

In \citet{Gong:2021tb}, the interloper fraction from the true redshift $j$ to the observed redshift bin $i$ is estimated by a simplistic ratio form, and we denote it as $f_0$ here.
\begin{equation}
    f_0 \equiv P_{ji}\approx\frac{C_{ji}^{gg,D}}{C_{jj}^{gg,D}}\ .
    \label{eqn:gongyan}
\end{equation}
The implicit assumption in the above estimation is that there are no interlopers for the observed redshift bin $j$.
However, this is largely not true in practice, especially when we chose a lot of redshift bins.
In more realistic case, the ratio, derived from equation~(\ref{eqn:CDsum}), is
\begin{equation}
    \frac{C_{ji}^{gg,D}}{C_{jj}^{gg,D}}=\frac{P_{jj}P_{ji}C_{jj}^{gg,R}}{P_{jj}P_{jj}C_{jj}^{gg,R}+P_{kj}P_{kj}C_{kk}^{gg,R}}\ .
    \label{eqn:fraction}
\end{equation}
Here, we further consider the interloper galaxies from the true redshift bin $k$ to the observed bin $j$.
For example, when using H$\alpha$ lines to determine the redshifts, redshift bins $i$, $j$ and $k$ can correspond to bins 4, 7 and 10, respectively.
Actually, we can derive the bias on $f_0$ from neglecting the secondary interlopers.
Denoting the interloper fraction $P_{kj}=f_1$, the equation~(\ref{eqn:fraction}) can be written by Taylor expansion as follows:
\begin{equation}
    \frac{C_{ji}^{gg,D}}{C_{jj}^{gg,D}}\approx f_0[1+f_1+(1-\frac{C_{kk}^{gg,R}}{C_{jj}^{gg,R}})f_1^2]\ .
    \label{eqn:fraction bias}
\end{equation}
We can see the ratio is close to $f_0$ only if the value of $f_1$ is very small.

In Fig.~\ref{fig:result_compare_gongyan}, We compare the results from the ratio method based on equation~(\ref{eqn:gongyan}).
We note that the improvement in accuracy of the ratio method relative to the results in \citet{Gong:2021tb} may be due to the use of angular power spectra instead of angular correlation function, or due to the difference in the scale range used in the analysis.
Besides, the error bars in \citet{Gong:2021tb} are derived by the average values of the errors in different angular bins.
Instead, here we estimate the error bars from the 10 groups.
It is obvious that in some redshift bins the reconstruction accuracy using the ratio method will degrade as the interloper fractions rises, which is in line with the expectation from equation~(\ref{eqn:fraction bias}).
However, the accuracy of the results obtained by our self-calibration method can be stable in different cases.

In fact, using the equation~(\ref{eqn:fraction bias}), we can derive an iterative method based on the relation between the ratio $C_{ji}^{gg,D}/C_{jj}^{gg,D}$ and $f_0$.
Because the redshifts of a given catalogue have the upper and lower limits.
For large $k$ or small $k$ in different cases, $P_{kk}=1$, and the interloper fraction $P_{kj}$ is exactly equal to the ratio $C_{kj}^{gg,D}/C_{kk}^{gg,D}$.
Then the estimated value of $P_{kj}$, i.e. $f_1$ in equation~(\ref{eqn:fraction bias}), can be used to obtain interloper fraction $f_0$.
We propose that, in this iterative way, all the true interloper fractions can be derived in order.
Here we simplify the equation~(\ref{eqn:fraction bias}) to
\begin{equation}
    \frac{C_{ji}^{gg,D}}{C_{jj}^{gg,D}}\approx f_0(1+f_1)\ ,
    \label{eqn:fraction bias simplified}
\end{equation}
because the interloper fractions are usually small.
Nevertheless, if the value of $C_{kk}^{gg,P}/C_{jj}^{gg,P}$ is found to be much larger than 1, we can use it to approximate the $C_{kk}^{gg,R}/C_{jj}^{gg,R}$ in equation~(\ref{eqn:fraction bias}) and do calculation.
The results after implementing this iteration method are shown in Fig.~\ref{fig:result_compare_method}.
We find it can successfully solve the problem in previous ratio method and the accuracy is comparable to the self-calibration algorithm.
However, when dealing with complex cases that require multi-step iterations, the accuracy of this approach may degrade.
Furthermore, as mentioned in \citet{Sun:2023aa}, this naive ratio estimator is sub-optimal and biased in terms of statistical errors.
Therefore, the self-calibration algorithm is more feasible and can be implemented in complicated cases.

\section{Interloper Contamination between Another Pair of Emission Lines}
\label{appendix_B}
\begin{figure*}
\centering
    \begin{minipage}{8.5cm}
        \centering
        \includegraphics[width=8.5cm]{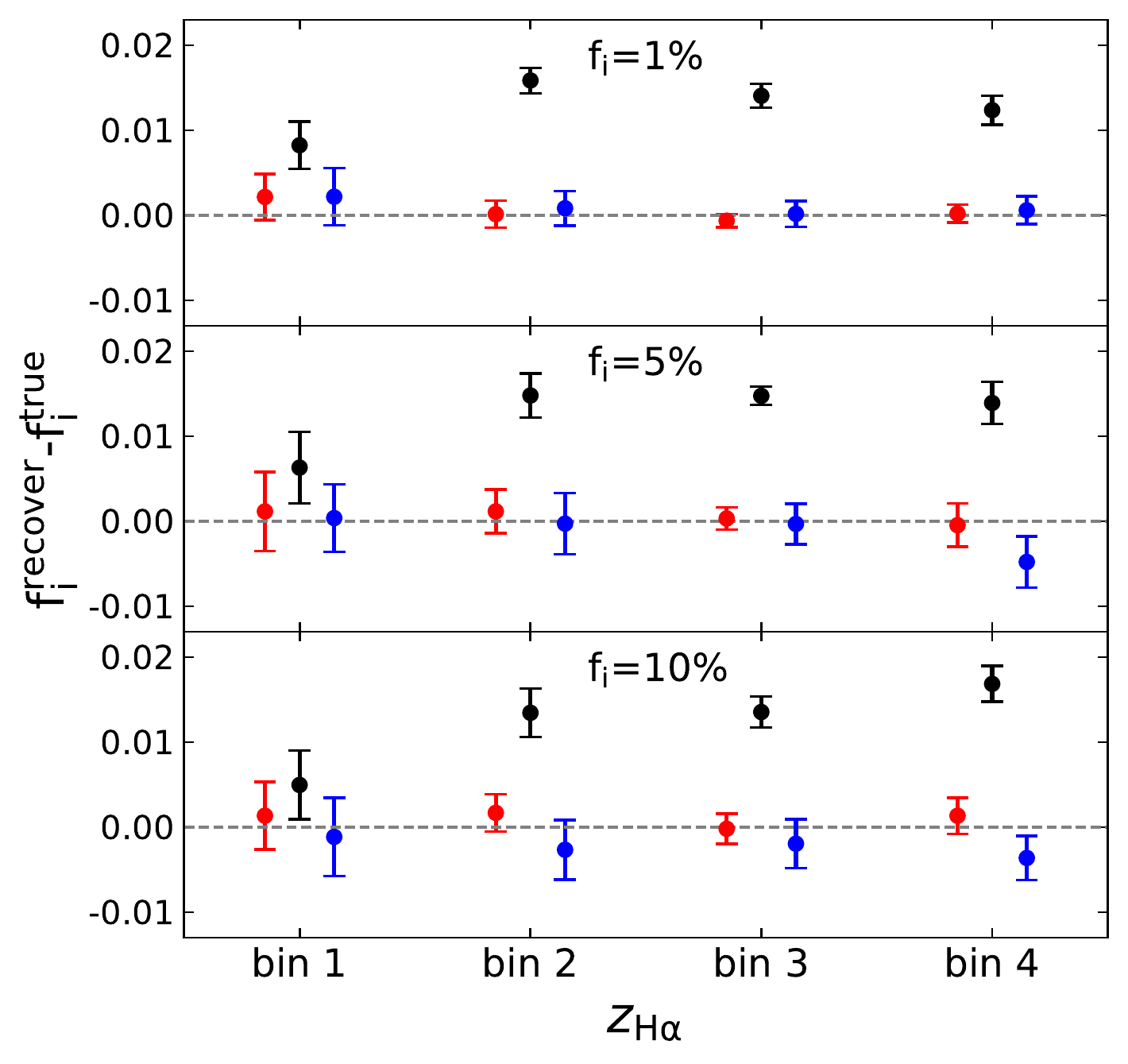}
    \end{minipage}
    \hspace{0.3cm}
	\begin{minipage}{8.5cm}
    	\centering
        \includegraphics[width=8.5cm]{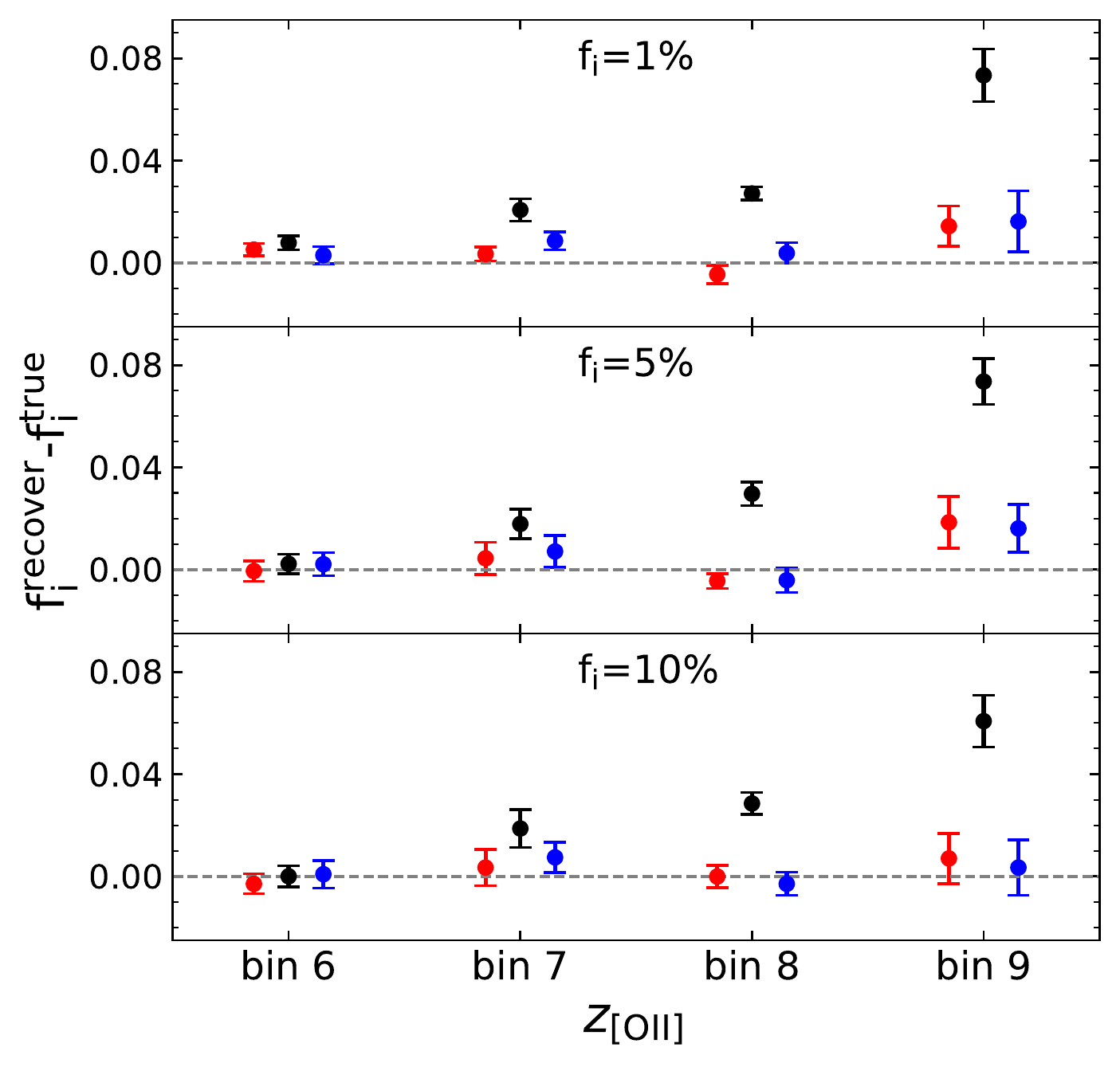}
    \end{minipage}
    \caption{Similar to Figs.~\ref{fig:result_interloper_mb} and \ref{fig:result_interloper_eliminate}, but adopting H$\alpha$ and [O\,\textsc{ii}] emission lines instead. The plot shows the reconstructed interloper fractions before lensing (red points), after lensing (black points) and implementing the elimination method (blue points).}
    \label{fig:result_Ha_OII}
\end{figure*}
Here we also investigate the accuracy of the self-calibration algorithm and elimination method on the contamination between H$\alpha$ 6563~{\AA} and [O\,\textsc{ii}] 3727~{\AA} emission lines, which are relatively further away from each other than the fiducial case.
To match the relation in equation~(\ref{eqn:interloper}), we divide the redshift range $0<z<1.478$ into another 9 tomographic bins with edges $z=$ 0.000, 0.100, 0.200, 0.300, 0.407, 0.761, 0.937, 1.113, 1.289, 1.478.
% The logarithmic luminosity slope $\alpha$ of 9 redshift bins are set to [0.20, 0.26, 0.35, 0.55, 1.20, 1.90, 2.35, 2.85, 3.50].
The contamination happens between bin $j$ and $j+5$, for $j=1, \cdots, 4$.

The results of adopting H$\alpha$ and [O\,\textsc{ii}] emission lines are shown in Fig.~\ref{fig:result_Ha_OII}, which is consistent with the results in the main text.
This consistence indicate that our method can generalize to any similar spectroscopic surveys.

% Don't change these lines
\bsp	% typesetting comment
\label{lastpage}
\end{document}